\documentclass[prl,superscriptaddress,amsmath,amssymb,twocolumn,nofootinbib]{revtex4-2}
\usepackage{graphicx,bm,amsmath}
\usepackage[usenames,dvipsnames]{xcolor}
\usepackage[colorlinks,bookmarks=false,citecolor=blue,linkcolor=Red,urlcolor=blue,
]{hyperref}
\usepackage[percent]{overpic}
\usepackage{bbm}

\usepackage[bbgreekl]{mathbbol}
\usepackage{xcolor}
\usepackage[normalem]{ulem}
\usepackage{braket}
\usepackage[ruled,vlined]{algorithm2e}
\usepackage{comment}
\usepackage{amsthm}
\usepackage{lipsum}
\usepackage{enumerate}
\usepackage{enumitem}
\usepackage{subfigure}
\graphicspath{{./figs/}}

\newcommand{\newsec}[1]{\textcolor{blue}{{\textit{#1}.-}}}

\newcommand{\Tr}{\text{Tr}}

\begin{document}
\title{Strong-to-Weak Spontaneous Symmetry Breaking \\in a $(2+1)$D Transverse-Field Ising Model under Decoherence}

\author{Yi-Ming Ding}
\affiliation{Department of Physics, School of Science and Research Center for Industries of the Future, Westlake University, Hangzhou 310030,  China}
\affiliation{Institute of Natural Sciences, Westlake Institute for Advanced Study, Hangzhou 310024, China}
\affiliation{State Key Laboratory of Surface Physics and Department of Physics, Fudan University, Shanghai 200438, China}

\author{Yuxuan Guo}
\email{yuxguo2024@g.ecc.u-tokyo.ac.jp}
\affiliation{Department of Physics, University of Tokyo, 7-3-1 Hongo, Bunkyo-ku, Tokyo 113-0033, Japan}

\author{Zhen Bi}
\email{zjb5184@psu.edu}
\affiliation{Department of Physics$,$ The Pennsylvania State University$,$ University Park$,$ Pennsylvania$,$ 16802$,$ USA}
\affiliation{Center for Theory of Emergent Quantum Matter$,$ Institute for Computational and Data Sciences$,$ The Pennsylvania State University$,$ University Park$,$ Pennsylvania$,$ 16802$,$ USA}

\author{Zheng Yan}
\email{zhengyan@westlake.edu.cn}
\affiliation{Department of Physics, School of Science and Research Center for Industries of the Future, Westlake University, Hangzhou 310030,  China}
\affiliation{Institute of Natural Sciences, Westlake Institute for Advanced Study, Hangzhou 310024, China}

\begin{abstract}
Decoherence in many-body quantum systems can give rise to intrinsically mixed-state phases and phase transitions beyond the pure-state paradigm. Here we study the $(2+1)$D transverse-field Ising model (TFIM) subject to a strongly $\mathbb{Z}_2$-symmetric decoherence channel, with a focus on strong-to-weak spontaneous symmetry breaking (SWSSB). This problem is challenging because the relevant transitions occur in the strong-decoherence regime, beyond the reach of perturbative expansions around the pure-state limit, while conventional quantum Monte Carlo (QMC) methods are hampered by the need to access nonlinear observables and by the sign problem. We overcome these difficulties by developing a QMC algorithm that efficiently evaluates nonlinear Rényi-2 correlators in higher dimensions, complemented by an effective field-theoretic approach. We show that the decohered state realizes a rich mixed-state phase diagram governed by an effective 2D Ashkin--Teller theory. This theory enables analytical predictions for the mixed-state phases and the universality classes of the phase boundaries, all of which are confirmed by large-scale QMC simulations.
\end{abstract}

\maketitle

\newsec{Introduction}
In recent years, rapid advances in quantum simulation platforms have brought nonequilibrium quantum matter to the forefront of condensed matter physics~\cite{Georgescu2014quantumsimulation,saffman2018quantumneutral,preskill2018nisq,zengb2019quantummatter}. In these systems, unavoidable coupling to the environment induces decoherence, which can qualitatively reshape collective behavior and give rise to novel mixed-state phases and phase transitions in interacting many-body systems~\cite{Lee:2022hog,leejy2023weakmeasurement,marc2023aspt,zouyj2023channeling,moharramipour2024symmetry,sukx2024higherform,PhysRevB.109.195420,yan2018interacting,fanrh2024diagnostics,sala2024swssbpurification,lessa2025swssb,zhang2025swssb1form,sunn2025swssb,PhysRevB.111.115123,guoyx2025qswssb,guoyc2025swssb,orito2025swssb_ising,guoyc2026purfication,sang2025markov,chenlxswssb,marc2025symmetry,marc2025topo,guoyc2025ldpo,ellison2025mixedtopo,lessa2025mixed-anomaly,ding2026mixedstatemipt,Sang2024mixedRG,Rakovszky2024defineOpenPhase,sang2025mixedstatephaseslocalreversibility,hasting2011finiteTtopo,wang2025anomalymixed,Hsin2024,Xu2025averageexacta,li2026generalizedsymmetryprotectedtopologicalphases,ziereis2025swssb,lucas2025swssbexact,wang2025decoherencetopo,Zhou2025finiteTotopo3D,putz2025learningtransitionsclassicalising}. A central question is how symmetry and symmetry breaking, a basic organizing principle of quantum phases of matter~\cite{Landau1937symmetry,Sachdev2011qpt,McGreevy2023symmetryreview}, generalize to open quantum systems. Here one must distinguish between strong and weak symmetry~\cite{Buca2012strongweak,albert2014symmetries,lieu2020strongweak}. A mixed state with strong symmetry is analogous to a canonical ensemble with fixed symmetry charge, whereas a mixed state with weak symmetry is analogous to a grand-canonical ensemble, in which different symmetry charges are mixed while the full density matrix remains symmetric. Accordingly, strong symmetry has a direct pure-state analog, whereas weak symmetry is meaningful only for mixed states. This distinction gives rise to strong-to-weak spontaneous symmetry breaking (SWSSB), in which strong symmetry is lost while weak symmetry remains~\cite{sala2024swssbpurification,leejy2023weakmeasurement,sang2025markov, Bao:2023zry,Ogunnaike:2023qyh,marc2023aspt,zhangjh2024fluctdissipat,fanrh2024diagnostics,kim2024errorthresholdsyk,lu2024bilayermixedstate,Gud2024swssb,kuno2024swssb_circuits_renyi2,lessa2025swssb,zhang2025swssb1form,sunn2025swssb,guoyx2025qswssb,guoyc2025swssb,orito2025swssb_ising,chenlxswssb,marc2025symmetry,huangxy2025hydro_SWSSB,luor2025topo-holo-mixed,liu2026fidelitympdo,zhang2025renyicorr_experi,zhangjh2025fidelitystrangecorrelator,songzj2025swssb,ziereis2025swssb,lucas2025swssbexact, feng2025random, hauser2026strongtoweaksymmetrybreakingopen, shu2026universaldynamicalscalingstrongtoweak, chen2025zippingmanybodyquantumstates}. It is a genuinely mixed-state form of order, with no counterpart in pure states.

Compared with conventional spontaneous symmetry breaking (SSB), diagnosing SWSSB requires nonlinear probes. Faithful order parameters include the fidelity and Rényi-1 correlators~\cite{lessa2025swssb,weinstein2025renyi1,liuzy2025wightman}, which obey the stability theorem of two-way connectivity via symmetric short-depth quantum channels, but are generally difficult to evaluate in many-body systems. By contrast, the Rényi-2 correlator, defined through a doubled pure-state representation via the Choi-Jamio\l{}kowski isomorphism~\cite{Choi1975,Jamiolkowski1972}, provides a practical proxy for detecting SWSSB~\cite{leejy2023weakmeasurement, sala2024swssbpurification}. 
Its advantage lies in both numerical and experimental accessibility: evaluating fidelity or Rényi-1 correlators requires canonical purification and thus full state tomography, whereas phase transitions of Rényi-2 observables can be efficiently measured using modern techniques such as classical shadow tomography~\cite{sunn2025swssb,Aaronson2019shadow,Huang2020shadow}.

Importantly, the Rényi-2 correlator is well suited for numerical approaches such as tensor-network methods based on matrix product states (MPS) in a doubled Hilbert space~\cite{guoyx2025qswssb,guoyc2025swssb,orito2025swssb_ising,guoyc2026purfication}.
However, these methods are largely restricted to one dimension and become challenging in higher dimensions. 
Quantum Monte Carlo (QMC) methods, by contrast, are naturally suited for large-scale simulations and largely independent of spatial dimension.  
Recent work has explored Monte Carlo sampling of SWSSB in specific models that admit classical stochastic representations~\cite{hauser2026swssb_u1}. 
However, a general QMC framework for directly computing the Rényi-2 correlator in interacting quantum systems, without relying on mapped classical representations, remains lacking. 
In this Letter, we develop a QMC framework for evaluating the Rényi-2 correlator, thereby filling this methodological gap and enabling unbiased large-scale simulations of decohered quantum states and SWSSB in arbitrary spatial dimensions.

As an illustrative example, we investigate the $(2+1)$D transverse-field Ising model (TFIM)  subject to a strongly $\mathbb{Z}_2$-symmetric decoherence channel. Despite its simplicity, the mixed-state phase diagram of this system remains largely unexplored except in the product-state limit~\cite{lessa2025swssb}. By combining a field-theoretical approach with our newly developed QMC techniques, we characterize the decohered Ising ground states across various parameter regimes. Our field-theoretical framework maps the effective defect action to the (1+1)D Ashkin-Teller model \cite{PhysRevB.24.5229}, predicting three distinct phases and identifying the universality classes of the phase transitions as well as the tricritical point. 
All these analytical predictions are firmly corroborated by large-scale simulations enabled by our QMC framework.

\newsec{Setup} 
We investigate the phase diagram of the decohered ground state of the TFIM as a function of Ising interaction and decoherence strength. Specifically, we consider a $L \times L$ 2D square lattice of spins with periodic boundary conditions described by the Hamiltonian  
$H = -J\sum_{\langle ij\rangle}Z_iZ_j - \sum_i X_i$, 
where $Z_i$ and $X_i$ are Pauli operators, and $\langle ij\rangle$ denotes the nearest-neighbor bonds. The quantum critical point is located at $J_c\approx 0.328474$~\citep{Henk2022qmc_ising}, which corresponds to a 3D Ising conformal field theory (CFT).
When $J<J_c$, the system is in a paramagnetic phase, and the ground state $\rho_0$ has a global strong $\mathbb{Z}_2$ symmetry under the symmetry operator $X\equiv \prod_i X_i$. When $J>J_c$, the strong $\mathbb{Z}_2$ symmetry is spontaneously broken to the trivial group, leading to a ferromagnetic order, as characterized by the linear order parameter $C^{(0)}\equiv\lim_{|i-j|\to \infty}\Tr(\rho_0 Z_iZ_j)=O(1)$.  

By applying the decoherence channel $\mathcal{E}=\prod_{\langle ij\rangle}\mathcal{E}_{\langle ij\rangle}$ to $\rho_0$, we obtain $\rho=\mathcal{E}[\rho_0]$, where each local channel preserves the strong $\mathbb{Z}_2$ symmetry and is defined as
\begin{equation}\label{eq:channel}
    \mathcal{E}_{\langle ij\rangle}[\rho_0] = \bigg(1-\frac{p}{2}\bigg) \rho_0 + \frac{p}{2} Z_iZ_j\rho_0 Z_i Z_j,\quad p\in[0, 1].
\end{equation}
We study the phase diagram of $\rho$ as a function of the decoherence strength $p$ and the coupling $J$. To do so, we first introduce several diagnostics for the distinct mixed-state phases. 

Under the Choi-Jamio\l{}kowski isomorphism~\cite{Choi1975,Jamiolkowski1972}, density matrix $\rho = \sum_{s,s'} \rho_{ss'} \ket{s}\!\bra{s'}$ is mapped onto a pure state $\left| \rho \right\rangle\!\rangle = \sum_{s,s'} \rho_{ss'} \ket{s}_{a} \ket{s'}^*_b$ in a doubled Hilbert space, where $a$ and $b$ label the two replicas. The symmetry of the system is given by $(\mathbb{Z}^a_2 \times \mathbb{Z}^b_2) \rtimes \mathbb{Z}_2^H$, where $\mathbb{Z}_2^{a/b}$ encode the strong symmetry in doubled space, while $\mathbb{Z}_2^H$ arises from Hermiticity and is strictly preserved. This symmetry structure allows two patterns of SSB \cite{leejy2023weakmeasurement}: complete breaking of $\mathbb{Z}_2^a \times \mathbb{Z}_2^b$, or partial breaking to the diagonal subgroup $\mathbb{Z}_2^{\mathrm{diag}}$, corresponding to SWSSB.
To diagnose the presence of SWSSB, we compute the Rényi-2 correlator $C^{(2)} \equiv \lim_{|i-j|\to\infty} \mathrm{Tr}(\rho Z_iZ_j\rho Z_iZ_j)/\mathrm{Tr}(\rho^2)$. 
A nonzero $C^{(2)}$ signals spontaneous breaking of the strong symmetry, but by itself does not diagnose the fate of the weak symmetry in the Choi doubled space.

To distinguish between the two symmetry-broken patterns, we also introduce the \emph{Rényi-2 linear order correlator in the doubled space}, defined as $C^{(1)} \equiv 
\lim_{|i-j|\to\infty}\mathrm{Tr}(\rho^2 Z_iZ_j) / \mathrm{Tr}(\rho^2)$.   
At the Choi-state level, $C^{(1)}=\lim_{|i-j|\to\infty}\langle\!\langle \rho| Z_i^{a/b}Z_j^{a/b}|\rho\rangle\!\rangle/\langle\!\langle \rho|\rho \rangle\!\rangle$ is the natural linear order parameter, which is nonzero only when the $\mathbb{Z}^a_2 \times \mathbb{Z}^b_2$ symmetry is completely broken, indicating the absence of weak symmetry of the original mixed state $\rho$. 

The three correlators $C^{(0)}$, $C^{(1)}$, and $C^{(2)}$ probe symmetry-breaking structure with increasing sensitivity to the Choi-state representation of $\rho$. Among them, $C^{(0)}$ diagnoses symmetry breaking directly at the level of the density matrix and therefore determines whether weak symmetry is broken. By contrast, $C^{(1)}$ and $C^{(2)}$ are nonlinear probes, sensitive to order that survives only in the doubled-state description. This clarifies the seemingly paradoxical regime $C^{(1)} \neq 0$ but $C^{(0)} = 0$: the density matrix itself remains weakly symmetric, while the associated Choi state still carries a finer symmetry-breaking structure. Compared with $C^{(0)}$, the correlator $C^{(1)}$ enhances the dominant contributions to $\rho$ and thus serves as a more sensitive probe of this hidden order. However, a nonzero $C^{(1)}$ should be understood as signaling symmetry breaking in the associated Choi doubled state, and not necessarily spontaneous symmetry breaking of the mixed state in the strict sense. We nevertheless refer to the regime with $C^{(1)} \neq 0$ and $C^{(0)} = 0$ as the \textit{Rényi-2 spontaneous symmetry breaking} (R2-SSB) regime. In contrast, when $C^{(2)} \neq 0$ but $C^{(1)} = 0$, the order is breaking $\mathbb{Z}_2^{a/b}$ down to $\mathbb{Z}_2^{\text{diag}}$ in the Choi state; we call this the \textit{Rényi-2 strong-to-weak spontaneous symmetry breaking} (R2-SWSSB) regime. Thus, the hierarchy of $C^{(0)}$, $C^{(1)}$, and $C^{(2)}$ distinguishes ordinary symmetry breaking of $\rho$ from progressively finer symmetry-breaking in its Choi-state representation.
A systematic review of these diagnostics is provided in the Supplemental Material~\cite{supplemental}.

\newsec{Quantum Monte Carlo method}
\begin{figure*}[ht!]
    \centering
    \includegraphics[width=\linewidth]{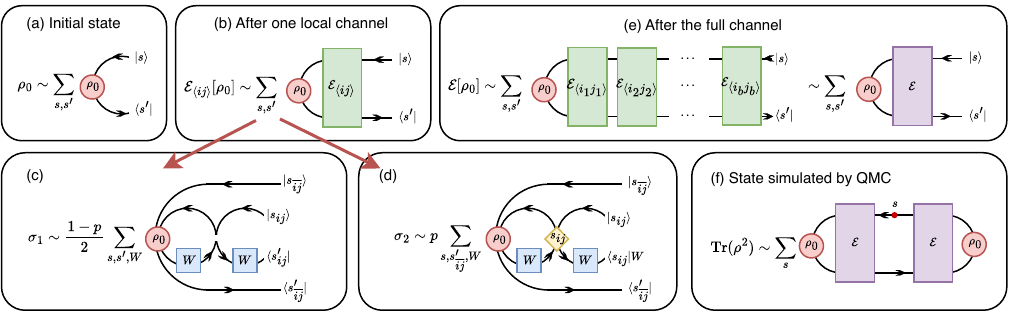}
    \caption{
    Graphical representation of the density matrix in the evolution picture, with propagation from the ket to the bra index in the computational basis.
    (a) Initial state $\rho_0$.
    (b) State after a single local channel.
    (c, d) Two contributions $\sigma_1$ and $\sigma_2$ in $\mathcal{E}_{\langle ij\rangle}[\rho_0]=\sigma_1+\sigma_2$ shown in (b).
    The yellow diamond in (d) denotes a Kronecker tensor enforcing identical spin states on sites $i$ and $j$ along four time directions.
    (e) State after the full channel.
    (f) Contracting two copies of $\rho=\mathcal{E}[\rho_0]$ yields $\rho^2$.
    Further contracting the bra and ket indices gives $\Tr(\rho^2)$, which can be simulated by QMC to evaluate Rényi-2 observables.
    }
    \label{fig:tensor}
\end{figure*}
Although the finite-temperature Gibbs state or ground state of a quantum Hamiltonian can be efficiently simulated using standard QMC methods based on imaginary-time path integral or series expansion ~\cite{prokof1998worm,sandvik1999sse,sandvik2003sseising,syljuasen2002directed,Boninsegni2006worm,Melko2013sse,Yan2019sweeping,yan2020improved,yan2022triangular}, directly implementing the local channel $\mathcal{E}_{\langle ij\rangle}$ is nontrivial.
In spin-$1/2$ models, the computational basis is typically chosen as the local $Z$ basis $\{\ket{s}\}$ with $s\in\{0,1\}^{\otimes N}$, where $\ket{0}\equiv\ket{\uparrow}$ and $\ket{1}\equiv\ket{\downarrow}$. 
The difficulty arises because operators such as $Z_i Z_j$ in Eq.~\eqref{eq:channel}, when acting within this computational basis, 
generically induce the notorious sign problem.

To circumvent this issue, we rewrite the local channel $\mathcal{E}_{\langle ij\rangle}$ in Eq.~\eqref{eq:channel} as $ \mathcal{E}_{\langle ij\rangle}[\rho_0]=\sum_k M_k \rho_0 M_k^\dagger$, 
where 
\begin{align}\label{eq:defm_k}
    M_0 &= \sqrt{1-p}\, \mathbbm{1}_i \mathbbm{1}_j , \nonumber\\
    M_1 &= \sqrt{p}\frac{\mathbbm{1}_i \mathbbm{1}_j+Z_iZ_j}{2},\quad M_2 = \sqrt{p}\frac{\mathbbm{1}_i \mathbbm{1}_j-Z_iZ_j}{2}.
\end{align} 
Then, the action of $\mathcal{E}_{\langle ij\rangle}$ on a general density matrix $ \rho_0=\sum_{s,s'} \langle s | \rho_0 | s' \rangle \ket{s}\bra{s'}$ can be written as 
$\mathcal{E}_{\langle ij\rangle}[\rho_0] = \sigma_1 + \sigma_2$, 
where 
\begin{align}
    \sigma_1 &\equiv \; \frac{1-p}{2}
    \sum_{s}\sum_{s'}\sum_W 
    \langle s |\rho_0 WW |s'\rangle
      \ket{s}\bra{s'}, 
      \label{eq:sigma1}
      \\ 
      \sigma_2  & \equiv \;  
      p
    \sum_{s}\sum_{s_{\overline{ij}}'}\sum_{W}
        \langle s|\rho_0 W  |s_{ij},s'_{\overline{ij}}\rangle \ket{s}\bra{s_{ij},s'_{\overline{ij}}}W,
        \label{eq:sigma2}
\end{align}
and $W\in\{ \mathbbm{1}_i\mathbbm{1}_j,\, X_i X_j \}$. 
For convenience, we decompose each basis state $\ket{s}=\otimes_k \ket{s_k}$ as $\ket{s} \equiv \ket{s_{ij}}\otimes \ket{s_{\overline{ij}}}$, where $\ket{s_{ij}} \equiv \ket{s_i}\otimes \ket{s_j}$ denotes the local degrees of freedom on sites $i$ and $j$, and $\ket{s_{\overline{ij}}}\equiv \otimes_{k\neq i,j}\ket{s_k}$ represents the remaining degrees of freedom. A more detailed derivation of Eqs.~\eqref{eq:sigma1} and \eqref{eq:sigma2} can be found in the Supplemental Material~\cite{supplemental}. 

The essential idea of QMC is to sample matrix elements of $\rho_0 \propto e^{-\beta H}$ with weights $W_{s,s'} \propto \langle s|\rho_0|s'\rangle$, without storing the full density matrix~\cite{mao2025sampling,wang2026generalized,mao2026detecting,ding2026mixedstatemipt}. As $\rho_0$ represents imaginary-time evolution [Fig.~\ref{fig:tensor}(a)], different matrix elements correspond to different temporal boundary conditions. In particular, sampling $\langle s|\rho_0|s\rangle$ imposes periodic boundary conditions by identifying the bra and ket states, yielding the standard partition function $Z=\Tr(e^{-\beta H})$. 
We note that matrix elements in $\sigma_1$ and $\sigma_2$ can be interpreted similarly [Fig.~\ref{fig:tensor}(b)-(d)]. 
Without loss of generality, we choose a counterclockwise time orientation. For $\sigma_2$, the resulting time contour for $s_{ij}$ has a figure-eight topology, and the orientation is taken along the outer loop.
The four-leg vertices in
Fig.~\ref{fig:tensor}(d) connect spin variables of sites $i$
and $j$ on different time branches and correspond to a rank-four
Kronecker tensor.
 
By iteratively applying the local channel $\mathcal{E}_{\langle ij\rangle}$, one obtains the decohered state $\rho = \mathcal{E}[\rho_0]$, whose matrix elements admit a similar graphical representation, as illustrated in Fig.~\ref{fig:tensor}(e).
In the same spirit, one may consider $\rho^2 = \bigl(\mathcal{E}[\rho_0]\bigr)^2$.
Imposing periodic boundary conditions along the time direction then allows the quantity $\Tr(\rho^2)$ to be interpreted as a generalized path integral and evaluated within QMC, as shown in Fig.~\ref{fig:tensor}(f).
Particularly, the quantities $C^{(1)}$ and $C^{(2)}$ for diagnosing SWSSB can be estimated using simple diagonal measurements by simulating $\Tr(\rho^2)$. Further details, including a discussion of the polynomial (and thus efficient) complexity of our algorithm, are provided in the Supplemental Material~\cite{supplemental}.

Finally, we emphasize that the QMC framework introduced here is general and applies to any initial state $\rho_0$ that can be prepared with standard QMC methods, including Gibbs states, ground states, and mixed states obtained by tracing out environmental degrees of freedom, and is not restricted to the TFIM considered here.

\begin{figure}[ht!]
    \centering
    \includegraphics[width=0.8\linewidth]{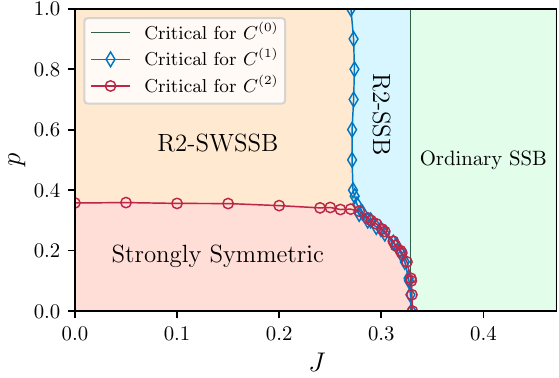}
    \caption{
    Phase diagram of the decohered ground state of the 2D TFIM under the quantum channel $\mathcal{E}$. Markers denote critical points obtained from QMC simulations associated with the corresponding order parameters. 
    For ground-state preparation, we take $\beta = 2L$, and verify convergence.
    The strongly symmetric phase is characterized by $C^{(\alpha)}=0$ for $\alpha=0,1,2$; 
the R2-SWSSB phase by $C^{(2)}\neq 0$ and $C^{(0)}=C^{(1)}=0$; 
the R2-SSB phase by $C^{(1)},C^{(2)}\neq 0$ with $C^{(0)}=0$; 
and the ordinary SSB phase by $C^{(\alpha)}\neq 0$ for all $\alpha=0,1,2$. 
    }
    \label{fig:phase_diagram}
\end{figure}

\newsec{Phase diagram and field theory analysis} Using the diagnostics introduced above, we map out the phase diagram of the model in the plane of Ising interaction and decoherence strength. As shown in Fig.~\ref{fig:phase_diagram}, the model exhibits a rich phase structure, including a strongly symmetric phase, an R2-SWSSB phase, an R2-SSB phase, and an ordinary SSB phase. 
We note that the ground-state ordered region $J>J_c$ is not destroyed by the decoherence considered here (see the discussion in the End Matter).
All phase boundaries are continuous transitions. In particular, the symmetric-to-R2-SWSSB boundary and the R2-SWSSB-to-R2-SSB boundary are both in the 2D Ising universality class. Moreover, the two phase transitions merge in part of the phase diagram into a critical line with continuously varying exponents.

To determine the phase boundaries, we consider the generalized Binder ratios~\cite{binder1981renorm}. For example, the Binder ratio associated with $C^{(2)}$ is defined as 
\begin{equation}
    R^{(2)}\equiv \frac{\sum_{i,j,k,l}\Tr(\rho Z_iZ_jZ_kZ_l\rho Z_iZ_jZ_kZ_l)}{[\sum_{i,j}\Tr(\rho Z_iZ_j\rho Z_iZ_j)]^2}.
\end{equation}
Crossings of $R^{(2)}$ for different system sizes locate the critical point. If the strong $\mathbb{Z}_2$ symmetry is preserved, $R^{(2)} \to 3$, whereas $R^{(2)} \to 1$ when it is broken. Similar ratios $R^{(\alpha)}$ can be defined for other $C^{(\alpha)}$ using the corresponding two- and four-point correlators. 

\begin{figure}[ht!]
    \centering
    \includegraphics[width=\linewidth]{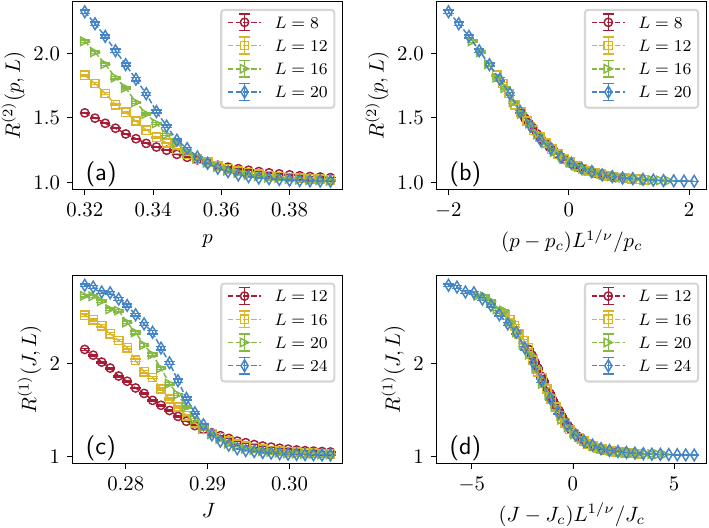}
    \caption{
    Binder ratios $R^{(\alpha)}$ as functions of system size $L$ and tuning parameters $J$ or $p$.
    (a) At fixed $J=0.1$, the curves cross at $p_c \approx 0.355$.
    (b) Corresponding data collapse of $R^{(2)}(p,L)$, yielding $\nu \approx 0.998\approx 1$.
    (c) At fixed $p=0.3$, the crossing occurs at $J_c \approx 0.29$, close to the tricritical point.
    (d) Data collapse using the expected exponent $\nu = 2/3$, with all curves collapsing near criticality, in agreement with theory.
    }
    \label{fig:plot_combine}
\end{figure}

We perform finite-size scaling to determine the universality classes of the phase boundaries in Fig.~\ref{fig:phase_diagram}. 
Before the red and blue boundaries merge, we find that the correlation-length exponent $\nu$ along both boundaries is consistent with the 2D Ising universality class. 
Fig.~\ref{fig:plot_combine}(a) shows an example at $J=0.1$, where tuning $p$ crosses the red critical boundary at $p_c \approx 0.355$. 
Finite-size scaling yields an excellent data collapse with $\nu \approx 0.998 \approx 1$ [Fig.~\ref{fig:plot_combine}(b)].
Similar behavior is observed for the other phase boundary, and a representative example is presented in the End Matter.
By contrast, as the two boundaries merge, the critical behavior changes qualitatively, with the correlation-length exponent deviating from the Ising value (as will be demonstrated below). To understand these results, we next develop an effective field-theoretical analysis.

One can formulate the problem of decoherence on the ground state of the TFIM in the path-integral formalism, where the ground state is prepared by imaginary-time evolution and the decoherence is encoded as an interaction along a codimension-one temporal defect~\cite{garratt2023, leejy2023weakmeasurement}. This leads to the effective action 
\begin{equation}
    S_{\text{eff}}[\phi_a,\phi_b] = S_{\text{Ising}}[\phi_a] + S_{\text{Ising}}[\phi_b] + S_{\text{int}}[\phi_a,\phi_b].
\end{equation}
The bulk corresponds to a standard $\phi^4$ action 
\begin{equation}
    S_{\text{Ising}}[\phi] = \frac{1}{2}\int d^2x dz \left( \partial_{\mu} \phi \partial^{\mu} \phi + m^2\phi^2 + \lambda\phi^4 \right),
\end{equation}
while the interaction induced by the decoherence only exists on a temporal defect:
\begin{equation}
    S_{\text{int}}[\phi_a,\phi_b] = -\frac{1}{2}\int d^2x dz \delta(z) \left[ \tilde{m}\left(\phi_a^2+\phi_b^2\right) + t\phi_a^2\phi_b^2 \right].
\end{equation}
This interaction arises from coarse-graining the microscopic interaction in the doubled space coming from the decoherence  $V_{\text{int}} = 2 \delta(z)\tanh^{-1}[p/(2-p)] \sum_{\langle i,j\rangle} Z_i^a Z_j^a Z_i^b Z_j^b$, implying $\tilde{m},t \propto \tanh^{-1}[p/(2-p)] $, see Supplemental Material~\cite{supplemental} for a detailed derivation. 

In the bulk ordered phase ($m^2<0$), the $\mathbb{Z}_2^a \times \mathbb{Z}_2^b$ symmetry is broken completely, and the resulting order is stable against symmetric finite-depth quantum channels. The system therefore lies in an ordinary SSB phase. In contrast, when the bulk is disordered ($m^2>0$), its finite correlation length implies that the bulk induces only short-range interactions on the defect, allowing for a much richer defect phase structure tuned by decoherence. The corresponding effective theory is equivalent to that of the 2D Ashkin-Teller model:
\begin{align}\label{eq:at_action}
    S_{\text{eff}}[\varphi_a,\varphi_b] &= \int d^2x \bigg[ \frac{1}{2}\sum_{\alpha=a,b} \Big( \partial_{\mu} \varphi^\alpha \partial^{\mu} \varphi^\alpha + m^2_{\text{eff}} \varphi_\alpha^2 \nonumber \\
    &\quad + \tilde{\lambda}\varphi_\alpha^4 \Big) - \tilde{t}\varphi_a^2\varphi_b^2 \bigg] + \cdots
\end{align}
where $m^2_{\text{eff}}=m(2m-\tilde{m})$, $\tilde{\lambda} = \lambda m / 2$ and $\tilde{t}=tm^2$.
We explicitly demonstrate this by integrating out the bulk fields in the path integral; detailed derivations are provided in the Supplemental Material~\cite{supplemental}. 

In the bulk disordered phase, the effective 2D Ashkin–Teller action~\eqref{eq:at_action} captures three distinct phases (see Supplemental Material~\cite{supplemental} for a Landau–Ginzburg analysis). 
For $m^2_{\text{eff}} > 0$ and ${t}$ is small compared to $\lambda/m$, no symmetry is broken, and the system is in a trivial symmetric phase. 
When $m^2_{\text{eff}} < 0$ and the inter-replica coupling $t$ is relatively small compared to $\lambda/m$, the Ashkin-Teller model enters the ferromagnetic phase (R2-SSB). This corresponds to the complete breaking of the $\mathbb{Z}^a_2 \times \mathbb{Z}^b_2$ symmetry to the trivial group. 
The third regime is the Baxter phase (R2-SWSSB), which emerges when $t$ is sufficiently large. 
In this phase, the $\mathbb{Z}^a_2 \times \mathbb{Z}^b_2$ symmetry is partially broken down to the diagonal subgroup $\mathbb{Z}_2^{\text{diag}}$. 
In particular, this field-theoretical analysis accounts for the meeting of these three phases at a tricritical point, in excellent agreement with the phase diagram in Fig.~\ref{fig:phase_diagram}.

Within the 2D Ashkin--Teller description, the transitions from the trivial to the Baxter phase (red boundary in Fig.~\ref{fig:phase_diagram}) and from the Baxter to the ferromagnetic phase (blue boundary) correspond to sequential breaking of two independent $\mathbb{Z}_2$ symmetries. Both transitions thus belong to the 2D Ising universality class, consistent with our numerical results.
More intriguingly, the direct transition between the strongly symmetric and R2-SSB phases (where the red and blue boundaries overlap in Fig.~\ref{fig:phase_diagram}) is governed by a compact boson CFT with a continuously tunable parameter, leading to continuously varying critical exponents along the phase boundary. Representative numerical results are presented in the End Matter.

Furthermore, the tricritical point where the strongly symmetric, R2-SWSSB, and R2-SSB phases meet is described by the 2D 4-state Potts CFT~\cite{DELFINO2004521}. The critical behavior is governed by the leading relevant energy-density operator $\varepsilon$ with scaling dimension $\Delta_\varepsilon = 1/2$, yielding $\nu = 1/(2-\Delta_\varepsilon) = 2/3$, while the subleading operator $\varepsilon'$ is marginal with $\Delta_{\varepsilon'}=2$. 
Although the tricritical point is difficult to locate precisely, data in its vicinity [Fig.~\ref{fig:plot_combine}(c), (d)] exhibit an excellent collapse with the expected exponent, supporting the Potts CFT description.

When $S_{\text{Ising}}$ is tuned to the critical point, the bulk correlation length diverges, which induces long-range interactions on the defect, invalidating the Ashkin-Teller description. Instead, this regime can be analyzed using conformal perturbation theory. Additional details and numerical confirmation are provided in the End Matter.

\newsec{Conclusions and outlook}
In this Letter, we propose a general QMC framework for evaluating nonlinear Rényi-2 correlators in decohered quantum many-body systems, filling a key methodological gap and enabling unbiased large-scale simulations of SWSSB in higher dimensions.
Combined with effective field-theoretical analysis, we systematically study the $(2+1)$D TFIM under a strongly $\mathbb{Z}_2$-symmetric decoherence channel, revealing a rich mixed-state phase diagram with three distinct phases: a strongly symmetric phase, a R2-SWSSB phase with partial symmetry breaking, and an R2-SSB phase with complete symmetry breaking in the doubled space.

Our field-theoretical approach provides a unified understanding of these phases and their transitions. 
Away from the Ising quantum critical point, the effective defect theory is governed by the 2D Ashkin--Teller model, leading to Ising-type transitions, a line of continuously varying criticality described by a compact boson CFT, and a tricritical point in the 2D 4-state Potts universality class. 
At the underlying quantum critical point, decoherence acts as a relevant defect perturbation, immediately driving the system into the R2-SSB phase with distinct critical scaling. 
These analytical predictions are in quantitative agreement with our QMC results.

Experimentally, both the TFIM and Pauli channels can be realized in quantum simulators~\cite{Simon2011tfim-exp,Labuhn2016tfim-exp,Ware2021pauli-exp,Lu2017pauli-exp,Chiuri2011pauli-exp}. Notably, SWSSB has recently been observed through Rényi-$k$ correlators using quantum gas microscopy~\cite{wang2026swssb-experiment}. Replica-based and randomized-measurement protocols also provide routes to the Rényi-2 diagnostics studied here~\cite{sunn2025swssb,Daley2012ee-replica-exp,Islam2015gas-micro-ee,Tiff2019random}.

Beyond the present model, our QMC and field theory framework opens several promising directions. It enables unbiased large-scale studies of mixed-state phases and phase transitions in a broad class of interacting systems, including higher dimensions, frustration, topological order, and continuous symmetries. It also naturally generalizes to a wide range of decoherence channels, including those preserving strong continuous symmetries and combinations of distinct channels. 
Moreover, combined with recent advances in QMC~\cite{melko2010ee,singh2011mutualinformation,DEmidio2020jazynski,ding2024reweight,DEmidio2024dqcp,Wang2025reweight,tarabunga2025bellqmc, ding2025negativity}, our framework can be extended to other nonlinear probes, such as entanglement entropy~\cite{zouyj2023channeling,ma2023exploringcriticalsystemsmeasurements}, entanglement Rényi negativity~\cite{fanrh2024diagnostics,caikl2026negativitytopodecohere,zouyj2023channeling,lu2024disentangling}, coherent information~\cite{fanrh2024diagnostics,Wang:2025awb}, and (conditional) mutual information~\cite{sang2025markov}, implemented via appropriate spacetime boundary conditions among replicas based on the graphical evolution picture introduced here.  These capabilities provide a powerful route to uncover and characterize new forms of mixed-state collective phenomena in the future. 

\newsec{Acknowledgments}
We thank Chong Wang, Yijian Zou, Yuto Ashida and Liujun Zou for helpful discussions. Y.G. is financially supported by the Global Science Graduate Course (GSGC) program at the University of Tokyo.  This project is supported by the Scientific Research Project (No. WU2025B011), Feng-Ying Career Development Chair Fund and the Start-up Fund of Westlake University. Z.B. acknowledges support from NSF CAREER Grant No.~DMR-2339319. The authors thank the IT service office and the high-performance computing center of Westlake University.


\newsec{Note added}
After completion of this work, we became aware of a parallel, independent, and complementary work by Z.~Weinstein and S.~J.~Garratt~\cite{weinstein2026toricthreshold}, which addresses the dual problem, the decoding problem in the $(2+1)$D transverse-field toric code under decoherence, using a distinct approach.

\newsec{Data availability}
The data supporting this work are publicly available on Zenodo~\cite{ding2026data}.

\bibliography{ref}

@article{Henk2022qmc_ising,
  title = {Cluster Monte Carlo simulation of the transverse Ising model},
  author = {Bl\"ote, Henk W. J. and Deng, Youjin},
  journal = {Phys. Rev. E},
  volume = {66},
  issue = {6},
  pages = {066110},
  numpages = {8},
  year = {2002},
  month = {Dec},
  publisher = {American Physical Society},
  doi = {10.1103/PhysRevE.66.066110},
  url = {https://link.aps.org/doi/10.1103/PhysRevE.66.066110}
}

@article{yan2018interacting,
  title = {Interacting lattice systems with quantum dissipation: A quantum Monte Carlo study},
  author = {Yan, Zheng and Pollet, Lode and Lou, Jie and Wang, Xiaoqun and Chen, Yan and Cai, Zi},
  journal = {Phys. Rev. B},
  volume = {97},
  issue = {3},
  pages = {035148},
  numpages = {7},
  year = {2018},
  month = {Jan},
  publisher = {American Physical Society},
  doi = {10.1103/PhysRevB.97.035148},
  url = {https://link.aps.org/doi/10.1103/PhysRevB.97.035148}
}

@article{syljuasen2002directed,
  title = {Quantum Monte Carlo with directed loops},
  author = {Sylju\aa{}sen, Olav F. and Sandvik, Anders W.},
  journal = {Phys. Rev. E},
  volume = {66},
  issue = {4},
  pages = {046701},
  numpages = {28},
  year = {2002},
  month = {Oct},
  publisher = {American Physical Society},
  doi = {10.1103/PhysRevE.66.046701},
  url = {https://link.aps.org/doi/10.1103/PhysRevE.66.046701}
}

@article{Yan2019sweeping,
  title = {Sweeping cluster algorithm for quantum spin systems with strong geometric restrictions},
  author = {Yan, Zheng and Wu, Yongzheng and Liu, Chenrong and Sylju\aa{}sen, Olav F. and Lou, Jie and Chen, Yan},
  journal = {Phys. Rev. B},
  volume = {99},
  issue = {16},
  pages = {165135},
  numpages = {6},
  year = {2019},
  month = {Apr},
  publisher = {American Physical Society},
  doi = {10.1103/PhysRevB.99.165135},
  url = {https://link.aps.org/doi/10.1103/PhysRevB.99.165135}
}

@article{mao2025sampling,
  title={Sampling reduced density matrix to extract fine levels of entanglement spectrum and restore entanglement Hamiltonian},
  author={Mao, Bin-Bin and Ding, Yi-Ming and Wang, Zhe and Hu, Shijie and Yan, Zheng},
  journal={Nature Communications},
  volume={16},
  number={1},
  pages={2880},
  year={2025},
  url={https://www.nature.com/articles/s41467-025-58058-0},
  publisher={Nature Publishing Group UK London}
}

@article{wang2026generalized,
  title={Generalized Reduced-Density-Matrix Quantum Monte Carlo Gives Access to More},
  author={Wang, Zhiyan and Wang, Zhe and Mao, Bin-Bin and Yan, Zheng},
  journal={arXiv preprint arXiv:2603.10948},
  year={2026}
}

@article{yan2022triangular,
  title={Triangular lattice quantum dimer model with variable dimer density},
  author={Yan, Zheng and Samajdar, Rhine and Wang, Yan-Cheng and Sachdev, Subir and Meng, Zi Yang},
  journal={Nature communications},
  volume={13},
  number={1},
  pages={5799},
  year={2022},
  url={https://www.nature.com/articles/s41467-022-33431-5},
  publisher={Nature Publishing Group UK London}
}

@article{prokof1998worm,
  title={“Worm” algorithm in quantum Monte Carlo simulations},
  author={Prokof'Ev, NV and Svistunov, BV},
  journal={Physics Letters A},
  volume={238},
  number={4-5},
  pages={253--257},
  year={1998},
  publisher={Elsevier}
}

@article{mao2026detecting,
  title={Detecting the Emergent Continuous Symmetry of Criticality via a Subsystem’s Entanglement Spectrum},
  author={Mao, Bin-Bin and Wang, Zhe and Chen, Bin-Bin and Yan, Zheng},
  journal={Physical Review Letters},
  volume={136},
  number={4},
  pages={046401},
  year={2026},
  url={https://journals.aps.org/prl/abstract/10.1103/7j21-l3pg},
  publisher={APS}
}

@article{Boninsegni2006worm,
  title = {Worm Algorithm for Continuous-Space Path Integral Monte Carlo Simulations},
  author = {Boninsegni, Massimo and Prokof'ev, Nikolay and Svistunov, Boris},
  journal = {Phys. Rev. Lett.},
  volume = {96},
  issue = {7},
  pages = {070601},
  numpages = {4},
  year = {2006},
  month = {Feb},
  publisher = {American Physical Society},
  doi = {10.1103/PhysRevLett.96.070601},
  url = {https://link.aps.org/doi/10.1103/PhysRevLett.96.070601}
}

@article{yan2020improved,
  title = {Global scheme of sweeping cluster algorithm to sample among topological sectors},
  author = {Yan, Zheng},
  journal = {Phys. Rev. B},
  volume = {105},
  issue = {18},
  pages = {184432},
  numpages = {9},
  year = {2022},
  month = {May},
  publisher = {American Physical Society},
  doi = {10.1103/PhysRevB.105.184432},
  url = {https://link.aps.org/doi/10.1103/PhysRevB.105.184432}
}

@Article{Huang2020shadow,
author={Huang, Hsin-Yuan
and Kueng, Richard
and Preskill, John},
title={Predicting many properties of a quantum system from very few measurements},
journal={Nature Physics},
year={2020},
month={Oct},
day={01},
volume={16},
number={10},
pages={1050-1057},
issn={1745-2481},
doi={10.1038/s41567-020-0932-7},
url={https://doi.org/10.1038/s41567-020-0932-7}
}

@inproceedings{Aaronson2019shadow,
author = {Aaronson, Scott and Rothblum, Guy N.},
title = {Gentle measurement of quantum states and differential privacy},
year = {2019},
isbn = {9781450367059},
publisher = {Association for Computing Machinery},
address = {New York, NY, USA},
url = {https://doi.org/10.1145/3313276.3316378},
doi = {10.1145/3313276.3316378},
booktitle = {Proceedings of the 51st Annual ACM SIGACT Symposium on Theory of Computing},
pages = {322–333},
numpages = {12},
location = {Phoenix, AZ, USA},
series = {STOC 2019}
}

@article{huangxy2025hydro_SWSSB,
  title = {Hydrodynamics as the effective field theory of strong-to-weak spontaneous symmetry breaking},
  author = {Huang, Xiaoyang and Qi, Marvin and Zhang, Jian-Hao and Lucas, Andrew},
  journal = {Phys. Rev. B},
  volume = {111},
  issue = {12},
  pages = {125147},
  numpages = {10},
  year = {2025},
  month = {Mar},
  publisher = {American Physical Society},
  doi = {10.1103/PhysRevB.111.125147},
  url = {https://link.aps.org/doi/10.1103/PhysRevB.111.125147}
}

@misc{hauser2026swssb_u1,
      title={Strong-to-Weak Symmetry Breaking in Open Quantum Systems: From Discrete Particles to Continuum Hydrodynamics}, 
      author={Jacob Hauser and Kaixiang Su and Hyunsoo Ha and Jerome Lloyd and Thomas G. Kiely and Romain Vasseur and Sarang Gopalakrishnan and Cenke Xu and Matthew P. A. Fisher},
      year={2026},
      eprint={2602.16045},
      archivePrefix={arXiv},
      primaryClass={quant-ph},
      url={https://arxiv.org/abs/2602.16045}, 
}

@article{guoyx2025qswssb,
  title = {Quantum Strong-To-Weak Spontaneous Symmetry Breaking in Decohered One-Dimensional Critical States},
  author = {Guo, Yuxuan and Yang, Sheng and Yu, Xue-Jia},
  journal = {PRX Quantum},
  volume = {6},
  issue = {4},
  pages = {040311},
  numpages = {21},
  year = {2025},
  month = {Oct},
  publisher = {American Physical Society},
  doi = {10.1103/4vs5-l54f},
  url = {https://link.aps.org/doi/10.1103/4vs5-l54f}
}

@article{kuno2024swssb_circuits_renyi2,
  title = {Strong-to-weak symmetry breaking states in stochastic dephasing stabilizer circuits},
  author = {Kuno, Yoshihito and Orito, Takahiro and Ichinose, Ikuo},
  journal = {Phys. Rev. B},
  volume = {110},
  issue = {9},
  pages = {094106},
  numpages = {12},
  year = {2024},
  month = {Sep},
  publisher = {American Physical Society},
  doi = {10.1103/PhysRevB.110.094106},
  url = {https://link.aps.org/doi/10.1103/PhysRevB.110.094106}
}

@misc{Gud2024swssb,
      title={Spontaneous symmetry breaking in open quantum systems: strong, weak, and strong-to-weak}, 
      author={Ding Gu and Zijian Wang and Zhong Wang},
      year={2024},
      eprint={2406.19381},
      archivePrefix={arXiv},
      primaryClass={quant-ph},
      url={https://arxiv.org/abs/2406.19381}, 
}

@article{guoyc2025swssb,
  title = {Strong-to-weak spontaneous symmetry breaking meets average symmetry-protected topological order},
  author = {Guo, Yuchen and Yang, Shuo},
  journal = {Phys. Rev. B},
  volume = {111},
  issue = {20},
  pages = {L201108},
  numpages = {6},
  year = {2025},
  month = {May},
  publisher = {American Physical Society},
  doi = {10.1103/PhysRevB.111.L201108},
  url = {https://link.aps.org/doi/10.1103/PhysRevB.111.L201108}
}

@misc{kim2024errorthresholdsyk,
      title={Error Threshold of SYK Codes from Strong-to-Weak Parity Symmetry Breaking}, 
      author={Jaewon Kim and Ehud Altman and Jong Yeon Lee},
      year={2024},
      eprint={2410.24225},
      archivePrefix={arXiv},
      primaryClass={quant-ph},
      url={https://arxiv.org/abs/2410.24225}, 
}

@misc{guoyc2026purfication,
      title={Quantum criticality in open quantum systems from the purification perspective}, 
      author={Yuchen Guo and Shuo Yang},
      year={2026},
      eprint={2602.21979},
      archivePrefix={arXiv},
      primaryClass={quant-ph},
      url={https://arxiv.org/abs/2602.21979}, 
}

@article{orito2025swssb_ising,
  title = {Strong and weak symmetries and their spontaneous symmetry breaking in mixed states emerging from the quantum Ising model under multiple decoherence},
  author = {Orito, Takahiro and Kuno, Yoshihito and Ichinose, Ikuo},
  journal = {Phys. Rev. B},
  volume = {111},
  issue = {5},
  pages = {054106},
  numpages = {11},
  year = {2025},
  month = {Feb},
  publisher = {American Physical Society},
  doi = {10.1103/PhysRevB.111.054106},
  url = {https://link.aps.org/doi/10.1103/PhysRevB.111.054106}
}

@Inbook{zengb2019quantummatter,
author="Zeng, Bei
and Chen, Xie
and Zhou, Duan-Lu
and Wen, Xiao-Gang",
bookTitle="Quantum Information Meets Quantum Matter: From Quantum Entanglement to Topological Phases of Many-Body Systems",
year="2019",
publisher="Springer New York",
address="New York, NY",
isbn="978-1-4939-9084-9",
doi="10.1007/978-1-4939-9084-9_1",
url="https://doi.org/10.1007/978-1-4939-9084-9_1"
}

@article{leejy2023weakmeasurement,
  title = {Quantum Criticality Under Decoherence or Weak Measurement},
  author = {Lee, Jong Yeon and Jian, Chao-Ming and Xu, Cenke},
  journal = {PRX Quantum},
  volume = {4},
  issue = {3},
  pages = {030317},
  numpages = {20},
  year = {2023},
  month = {Aug},
  publisher = {American Physical Society},
  doi = {10.1103/PRXQuantum.4.030317},
  url = {https://link.aps.org/doi/10.1103/PRXQuantum.4.030317}
}

@article{moharramipour2024symmetry,
  title = {Symmetry-Enforced Entanglement in Maximally Mixed States},
  author = {Moharramipour, Amin and Lessa, Leonardo A. and Wang, Chong and Hsieh, Timothy H. and Sahu, Subhayan},
  journal = {PRX Quantum},
  volume = {5},
  issue = {4},
  pages = {040336},
  numpages = {26},
  year = {2024},
  month = {Dec},
  publisher = {American Physical Society},
  doi = {10.1103/PRXQuantum.5.040336},
  url = {https://link.aps.org/doi/10.1103/PRXQuantum.5.040336}
}

@article{sukx2024higherform,
  title = {Higher-Form Symmetries under Weak Measurement},
  author = {Su, Kaixiang and Myerson-Jain, Nayan and Wang, Chong and Jian, Chao-Ming and Xu, Cenke},
  journal = {Phys. Rev. Lett.},
  volume = {132},
  issue = {20},
  pages = {200402},
  numpages = {6},
  year = {2024},
  month = {May},
  publisher = {American Physical Society},
  doi = {10.1103/PhysRevLett.132.200402},
  url = {https://link.aps.org/doi/10.1103/PhysRevLett.132.200402}
}

@article{fanrh2024diagnostics,
  title = {Diagnostics of Mixed-State Topological Order and Breakdown of Quantum Memory},
  author = {Fan, Ruihua and Bao, Yimu and Altman, Ehud and Vishwanath, Ashvin},
  journal = {PRX Quantum},
  volume = {5},
  issue = {2},
  pages = {020343},
  numpages = {17},
  year = {2024},
  month = {May},
  publisher = {American Physical Society},
  doi = {10.1103/PRXQuantum.5.020343},
  url = {https://link.aps.org/doi/10.1103/PRXQuantum.5.020343}
}

@article{zouyj2023channeling,
  title = {Channeling Quantum Criticality},
  author = {Zou, Yijian and Sang, Shengqi and Hsieh, Timothy H.},
  journal = {Phys. Rev. Lett.},
  volume = {130},
  issue = {25},
  pages = {250403},
  numpages = {7},
  year = {2023},
  month = {Jun},
  publisher = {American Physical Society},
  doi = {10.1103/PhysRevLett.130.250403},
  url = {https://link.aps.org/doi/10.1103/PhysRevLett.130.250403}
}

@misc{putz2025learningtransitionsclassicalising,
      title={Learning transitions in classical Ising models and deformed toric codes}, 
      author={Malte Pütz and Samuel J. Garratt and Hidetoshi Nishimori and Simon Trebst and Guo-Yi Zhu},
      year={2025},
      eprint={2504.12385},
      archivePrefix={arXiv},
      primaryClass={cond-mat.stat-mech},
      url={https://arxiv.org/abs/2504.12385}, 
}

@misc{weinstein2026toricthreshold,
      title={Intrinsic Error Thresholds in Nearly Critical Toric Codes}, 
      author={Zack Weinstein and Samuel J. Garratt},
      year={2026},
      eprint={2603.14098},
      archivePrefix={arXiv},
      primaryClass={cond-mat.stat-mech},
      url={https://arxiv.org/abs/2603.14098}, 
}

@article{lessa2025mixed-anomaly,
  title = {Mixed-State Quantum Anomaly and Multipartite Entanglement},
  author = {Lessa, Leonardo A. and Cheng, Meng and Wang, Chong},
  journal = {Phys. Rev. X},
  volume = {15},
  issue = {1},
  pages = {011069},
  numpages = {26},
  year = {2025},
  month = {Mar},
  publisher = {American Physical Society},
  doi = {10.1103/PhysRevX.15.011069},
  url = {https://link.aps.org/doi/10.1103/PhysRevX.15.011069}
}

@dataset{ding2026data,
  author       = {Ding, Yi-Ming},
  title        = {Data for "Strong-to-Weak Spontaneous Symmetry
                   Breaking in a (2+1)D Transverse-Field Ising Model
                   under Decoherence"
                  },
  month        = apr,
  year         = 2026,
  publisher    = {Zenodo},
  doi          = {10.5281/zenodo.19363042},
  url          = {https://doi.org/10.5281/zenodo.19363042},
}

@article{ellison2025mixedtopo,
  title = {Toward a Classification of Mixed-State Topological Orders in Two Dimensions},
  author = {Ellison, Tyler D. and Cheng, Meng},
  journal = {PRX Quantum},
  volume = {6},
  issue = {1},
  pages = {010315},
  numpages = {44},
  year = {2025},
  month = {Jan},
  publisher = {American Physical Society},
  doi = {10.1103/PRXQuantum.6.010315},
  url = {https://link.aps.org/doi/10.1103/PRXQuantum.6.010315}
}

@misc{caikl2026negativitytopodecohere,
      title={Entanglement negativity in decohered topological states}, 
      author={Kang-Le Cai and Meng Cheng},
      year={2026},
      eprint={2602.16597},
      archivePrefix={arXiv},
      primaryClass={cond-mat.str-el},
      url={https://arxiv.org/abs/2602.16597}, 
}

@misc{songzj2025swssb,
      title={Strong-to-weak spontaneous symmetry breaking of higher-form non-invertible symmetries in Kitaev's quantum double model}, 
      author={Zijian Song and Jian-Hao Zhang},
      year={2025},
      eprint={2509.24179},
      archivePrefix={arXiv},
      primaryClass={quant-ph},
      url={https://arxiv.org/abs/2509.24179}, 
}

@misc{zhangjh2025fidelitystrangecorrelator,
      title={Fidelity Strange Correlators for Average Symmetry-Protected Topological Phases}, 
      author={Jian-Hao Zhang and Yang Qi and Zhen Bi},
      year={2025},
      eprint={2210.17485},
      archivePrefix={arXiv},
      primaryClass={cond-mat.str-el},
      url={https://arxiv.org/abs/2210.17485}, 
}

@misc{zhangjh2024fluctdissipat,
      title={Fluctuation-Dissipation Theorem and Information Geometry in Open Quantum Systems}, 
      author={Jian-Hao Zhang and Cenke Xu and Yichen Xu},
      year={2024},
      eprint={2409.18944},
      archivePrefix={arXiv},
      primaryClass={quant-ph},
      url={https://arxiv.org/abs/2409.18944}, 
}

@article{lessa2025swssb,
  title = {Strong-to-Weak Spontaneous Symmetry Breaking in Mixed Quantum States},
  author = {Lessa, Leonardo A. and Ma, Ruochen and Zhang, Jian-Hao and Bi, Zhen and Cheng, Meng and Wang, Chong},
  journal = {PRX Quantum},
  volume = {6},
  issue = {1},
  pages = {010344},
  numpages = {24},
  year = {2025},
  month = {Mar},
  publisher = {American Physical Society},
  doi = {10.1103/PRXQuantum.6.010344},
  url = {https://link.aps.org/doi/10.1103/PRXQuantum.6.010344}
}

@article{zhang2025swssb1form,
  title = {Strong-to-weak spontaneous breaking of 1-form symmetry and intrinsically mixed topological order},
  author = {Zhang, Carolyn and Xu, Yichen and Zhang, Jian-Hao and Xu, Cenke and Bi, Zhen and Luo, Zhu-Xi},
  journal = {Phys. Rev. B},
  volume = {111},
  issue = {11},
  pages = {115137},
  numpages = {28},
  year = {2025},
  month = {Mar},
  publisher = {American Physical Society},
  doi = {10.1103/PhysRevB.111.115137},
  url = {https://link.aps.org/doi/10.1103/PhysRevB.111.115137}
}

@misc{liu2026fidelitympdo,
      title={Polynomial-time certification of fidelity for many-body mixed states and mixed-state universality classes}, 
      author={Yuhan Liu and Yijian Zou},
      year={2026},
      eprint={2601.13333},
      archivePrefix={arXiv},
      primaryClass={quant-ph},
      url={https://arxiv.org/abs/2601.13333}, 
}

@misc{zhang2025renyicorr_experi,
      title={Probing mixed-state phases on a quantum computer via Renyi correlators and variational decoding}, 
      author={Yuxuan Zhang and Timothy H. Hsieh and Yong Baek Kim and Yijian Zou},
      year={2025},
      eprint={2505.02900},
      archivePrefix={arXiv},
      primaryClass={quant-ph},
      url={https://arxiv.org/abs/2505.02900}, 
}

@article{albert2014symmetries,
  title = {Symmetries and conserved quantities in Lindblad master equations},
  author = {Albert, Victor V. and Jiang, Liang},
  journal = {Phys. Rev. A},
  volume = {89},
  issue = {2},
  pages = {022118},
  numpages = {14},
  year = {2014},
  month = {Feb},
  publisher = {American Physical Society},
  doi = {10.1103/PhysRevA.89.022118},
  url = {https://link.aps.org/doi/10.1103/PhysRevA.89.022118}
}

@misc{ding2026mixedstatemipt,
      title={Mixed-State Measurement-Induced Phase Transitions in Imaginary-Time Dynamics}, 
      author={Yi-Ming Ding and Zenan Liu and Xu Tian and Zhe Wang and Yanzhang Zhu and Zheng Yan},
      year={2026},
      eprint={2511.04402},
      archivePrefix={arXiv},
      primaryClass={quant-ph},
      url={https://arxiv.org/abs/2511.04402}, 
}

@article{lu2024disentangling,
  title = {Disentangling transitions in topological order induced by boundary decoherence},
  author = {Lu, Tsung-Cheng},
  journal = {Phys. Rev. B},
  volume = {110},
  issue = {12},
  pages = {125145},
  numpages = {6},
  year = {2024},
  month = {Sep},
  publisher = {American Physical Society},
  doi = {10.1103/PhysRevB.110.125145},
  url = {https://link.aps.org/doi/10.1103/PhysRevB.110.125145}
}

@misc{ma2023exploringcriticalsystemsmeasurements,
      title={Exploring critical systems under measurements and decoherence via Keldysh field theory}, 
      author={Ruochen Ma},
      year={2023},
      eprint={2304.08277},
      archivePrefix={arXiv},
      primaryClass={quant-ph},
      url={https://arxiv.org/abs/2304.08277}, 
}

@article{sunn2025swssb,
  title = {Scheme to Detect the Strong-to-Weak Symmetry Breaking via Randomized Measurements},
  author = {Sun, Ning and Zhang, Pengfei and Feng, Lei},
  journal = {Phys. Rev. Lett.},
  volume = {135},
  issue = {9},
  pages = {090403},
  numpages = {7},
  year = {2025},
  month = {Aug},
  publisher = {American Physical Society},
  doi = {10.1103/7p5x-7yqb},
  url = {https://link.aps.org/doi/10.1103/7p5x-7yqb}
}

@article{weinstein2025renyi1,
  title = {Efficient Detection of Strong-to-Weak Spontaneous Symmetry Breaking via the R\'enyi-1 Correlator},
  author = {Weinstein, Zack},
  journal = {Phys. Rev. Lett.},
  volume = {134},
  issue = {15},
  pages = {150405},
  numpages = {7},
  year = {2025},
  month = {Apr},
  publisher = {American Physical Society},
  doi = {10.1103/PhysRevLett.134.150405},
  url = {https://link.aps.org/doi/10.1103/PhysRevLett.134.150405}
}

@article{Choi1975,
title = {Completely positive linear maps on complex matrices},
journal = {Linear Algebra and its Applications},
volume = {10},
number = {3},
pages = {285-290},
year = {1975},
issn = {0024-3795},
doi = {https://doi.org/10.1016/0024-3795(75)90075-0},
url = {https://www.sciencedirect.com/science/article/pii/0024379575900750},
author = {Man-Duen Choi}
}

@article{Jamiolkowski1972,
title = {Linear transformations which preserve trace and positive semidefiniteness of operators},
journal = {Reports on Mathematical Physics},
volume = {3},
number = {4},
pages = {275-278},
year = {1972},
issn = {0034-4877},
doi = {https://doi.org/10.1016/0034-4877(72)90011-0},
url = {https://www.sciencedirect.com/science/article/pii/0034487772900110},
author = {A. Jamiołkowski}
}

@article{marc2025topo,
  title = {Topological Phases with Average Symmetries: The Decohered, the Disordered, and the Intrinsic},
  author = {Ma, Ruochen and Zhang, Jian-Hao and Bi, Zhen and Cheng, Meng and Wang, Chong},
  journal = {Phys. Rev. X},
  volume = {15},
  issue = {2},
  pages = {021062},
  numpages = {33},
  year = {2025},
  month = {May},
  publisher = {American Physical Society},
  doi = {10.1103/PhysRevX.15.021062},
  url = {https://link.aps.org/doi/10.1103/PhysRevX.15.021062}
}

@article{marc2023aspt,
  title = {Average Symmetry-Protected Topological Phases},
  author = {Ma, Ruochen and Wang, Chong},
  journal = {Phys. Rev. X},
  volume = {13},
  issue = {3},
  pages = {031016},
  numpages = {24},
  year = {2023},
  month = {Aug},
  publisher = {American Physical Society},
  doi = {10.1103/PhysRevX.13.031016},
  url = {https://link.aps.org/doi/10.1103/PhysRevX.13.031016}
}

@article{marc2025symmetry,
  title = {Symmetry-Protected Topological Phases of Mixed States in the Doubled Space},
  author = {Ma, Ruochen and Turzillo, Alex},
  journal = {PRX Quantum},
  volume = {6},
  issue = {1},
  pages = {010348},
  numpages = {33},
  year = {2025},
  month = {Mar},
  publisher = {American Physical Society},
  doi = {10.1103/PRXQuantum.6.010348},
  url = {https://link.aps.org/doi/10.1103/PRXQuantum.6.010348}
}

@misc{lu2024bilayermixedstate,
      title={Bilayer construction for mixed state phenomena with strong, weak symmetries and symmetry breakings}, 
      author={Shuangyuan Lu and Penghao Zhu and Yuan-Ming Lu},
      year={2024},
      eprint={2411.07174},
      archivePrefix={arXiv},
      primaryClass={cond-mat.str-el},
      url={https://arxiv.org/abs/2411.07174}, 
}

@article{guoyc2025ldpo,
  title = {Locally Purified Density Operators for Symmetry-Protected Topological Phases in Mixed States},
  author = {Guo, Yuchen and Zhang, Jian-Hao and Zhang, Hao-Ran and Yang, Shuo and Bi, Zhen},
  journal = {Phys. Rev. X},
  volume = {15},
  issue = {2},
  pages = {021060},
  numpages = {36},
  year = {2025},
  month = {May},
  publisher = {American Physical Society},
  doi = {10.1103/PhysRevX.15.021060},
  url = {https://link.aps.org/doi/10.1103/PhysRevX.15.021060}
}

@article{luor2025topo-holo-mixed,
  title = {Topological Holography for Mixed-State Phases and Phase Transitions},
  author = {Luo, Ran and Wang, Yi-Nan and Bi, Zhen},
  journal = {PRX Quantum},
  volume = {6},
  issue = {4},
  pages = {040358},
  numpages = {38},
  year = {2025},
  month = {Dec},
  publisher = {American Physical Society},
  doi = {10.1103/9kmh-gjf8},
  url = {https://link.aps.org/doi/10.1103/9kmh-gjf8}
}

@article{sala2024swssbpurification,
  title = {Spontaneous strong symmetry breaking in open systems: Purification perspective},
  author = {Sala, Pablo and Gopalakrishnan, Sarang and Oshikawa, Masaki and You, Yizhi},
  journal = {Phys. Rev. B},
  volume = {110},
  issue = {15},
  pages = {155150},
  numpages = {28},
  year = {2024},
  month = {Oct},
  publisher = {American Physical Society},
  doi = {10.1103/PhysRevB.110.155150},
  url = {https://link.aps.org/doi/10.1103/PhysRevB.110.155150}
}

@article{chenlxswssb,
  title = {Strong-to-weak symmetry breaking and entanglement transitions},
  author = {Chen, Langxuan and Sun, Ning and Zhang, Pengfei},
  journal = {Phys. Rev. B},
  volume = {111},
  issue = {6},
  pages = {L060304},
  numpages = {6},
  year = {2025},
  month = {Feb},
  publisher = {American Physical Society},
  doi = {10.1103/PhysRevB.111.L060304},
  url = {https://link.aps.org/doi/10.1103/PhysRevB.111.L060304}
}

@Article{liuzy2025wightman,
author={Liu, Zeyu
and Chen, Langxuan
and Zhang, Yuke
and Zhou, Shuyan
and Zhang, Pengfei},
title={Diagnosing strong-to-weak symmetry breaking via Wightman correlators},
journal={Communications Physics},
year={2025},
month={Jul},
day={02},
volume={8},
number={1},
pages={274},
issn={2399-3650},
doi={10.1038/s42005-025-02199-7},
url={https://doi.org/10.1038/s42005-025-02199-7}
}

@article{Georgescu2014quantumsimulation,
  title = {Quantum simulation},
  author = {Georgescu, I. M. and Ashhab, S. and Nori, Franco},
  journal = {Rev. Mod. Phys.},
  volume = {86},
  issue = {1},
  pages = {153--185},
  numpages = {33},
  year = {2014},
  month = {Mar},
  publisher = {American Physical Society},
  doi = {10.1103/RevModPhys.86.153},
  url = {https://link.aps.org/doi/10.1103/RevModPhys.86.153}
}

@article{saffman2018quantumneutral,
    author = {Saffman, Mark},
    title = {Quantum computing with neutral atoms},
    journal = {National Science Review},
    volume = {6},
    number = {1},
    pages = {24-25},
    year = {2018},
    month = {09},
    issn = {2095-5138},
    doi = {10.1093/nsr/nwy088},
    url = {https://doi.org/10.1093/nsr/nwy088},
    eprint = {https://academic.oup.com/nsr/article-pdf/6/1/24/38915067/nwy088.pdf},
}

@article{Landau1937symmetry,
      author        = "Landau, Lev Davidovich",
      title         = "{On the theory of phase transitions. I.}",
      journal       = "Phys. Z. Sowjet.",
      volume        = "11",
      pages         = "26",
      year          = "1937",
      url           = "https://cds.cern.ch/record/480039",
}

@article{Buca2012strongweak,
doi = {10.1088/1367-2630/14/7/073007},
url = {https://doi.org/10.1088/1367-2630/14/7/073007},
year = {2012},
month = {jul},
publisher = {IOP Publishing},
volume = {14},
number = {7},
pages = {073007},
author = {Buča, Berislav and Prosen, Tomaž},
title = {A note on symmetry reductions of the Lindblad equation: transport in constrained open spin chains},
journal = {New Journal of Physics}
}

@article{lieu2020strongweak,
  title = {Symmetry Breaking and Error Correction in Open Quantum Systems},
  author = {Lieu, Simon and Belyansky, Ron and Young, Jeremy T. and Lundgren, Rex and Albert, Victor V. and Gorshkov, Alexey V.},
  journal = {Phys. Rev. Lett.},
  volume = {125},
  issue = {24},
  pages = {240405},
  numpages = {7},
  year = {2020},
  month = {Dec},
  publisher = {American Physical Society},
  doi = {10.1103/PhysRevLett.125.240405},
  url = {https://link.aps.org/doi/10.1103/PhysRevLett.125.240405}
}

@article{sang2025markov,
  title = {Stability of Mixed-State Quantum Phases via Finite Markov Length},
  author = {Sang, Shengqi and Hsieh, Timothy H.},
  journal = {Phys. Rev. Lett.},
  volume = {134},
  issue = {7},
  pages = {070403},
  numpages = {7},
  year = {2025},
  month = {Feb},
  publisher = {American Physical Society},
  doi = {10.1103/PhysRevLett.134.070403},
  url = {https://link.aps.org/doi/10.1103/PhysRevLett.134.070403}
}

@book{Sachdev2011qpt, 
place={Cambridge}, 
edition={2}, 
title={Quantum Phase Transitions}, 
publisher={Cambridge University Press}, 
author={Sachdev, Subir}, 
year={2011}}

@article{preskill2018nisq,
   title={Quantum Computing in the NISQ era and beyond},
   volume={2},
   ISSN={2521-327X},
   url={http://dx.doi.org/10.22331/q-2018-08-06-79},
   DOI={10.22331/q-2018-08-06-79},
   journal={Quantum},
   publisher={Verein zur Forderung des Open Access Publizierens in den Quantenwissenschaften},
   author={Preskill, John},
   year={2018},
   month=aug, pages={79} }

@article{sandvik2003sseising,
  title = {Stochastic series expansion method for quantum Ising models with arbitrary interactions},
  author = {Sandvik, Anders W.},
  journal = {Phys. Rev. E},
  volume = {68},
  issue = {5},
  pages = {056701},
  numpages = {9},
  year = {2003},
  month = {Nov},
  publisher = {American Physical Society},
  doi = {10.1103/PhysRevE.68.056701},
  url = {https://link.aps.org/doi/10.1103/PhysRevE.68.056701}
}

@article{Rakovszky2024defineOpenPhase,
  title = {Defining Stable Phases of Open Quantum Systems},
  author = {Rakovszky, Tibor and Gopalakrishnan, Sarang and von Keyserlingk, Curt},
  journal = {Phys. Rev. X},
  volume = {14},
  issue = {4},
  pages = {041031},
  numpages = {35},
  year = {2024},
  month = {Nov},
  publisher = {American Physical Society},
  doi = {10.1103/PhysRevX.14.041031},
  url = {https://link.aps.org/doi/10.1103/PhysRevX.14.041031}
}

@misc{ziereis2025swssb,
      title={Strong-to-Weak Symmetry Breaking Phases in Steady States of Quantum Operations}, 
      author={Niklas Ziereis and Sanjay Moudgalya and Michael Knap},
      year={2025},
      eprint={2509.09669},
      archivePrefix={arXiv},
      primaryClass={cond-mat.stat-mech},
      url={https://arxiv.org/abs/2509.09669}, 
}

@article{Zhou2025finiteTotopo3D,
  title = {Finite-Temperature Quantum Topological Order in Three Dimensions},
  author = {Zhou, Shu-Tong and Cheng, Meng and Rakovszky, Tibor and von Keyserlingk, Curt and Ellison, Tyler D.},
  journal = {Phys. Rev. Lett.},
  volume = {135},
  issue = {4},
  pages = {040402},
  numpages = {7},
  year = {2025},
  month = {Jul},
  publisher = {American Physical Society},
  doi = {10.1103/n9sq-8cxw},
  url = {https://link.aps.org/doi/10.1103/n9sq-8cxw}
}

@misc{wang2025decoherencetopo,
      title={Decoherence-induced self-dual criticality in topological states of matter}, 
      author={Qingyuan Wang and Romain Vasseur and Simon Trebst and Andreas W. W. Ludwig and Guo-Yi Zhu},
      year={2025},
      eprint={2502.14034},
      archivePrefix={arXiv},
      primaryClass={quant-ph},
      url={https://arxiv.org/abs/2502.14034}, 
}

@article{lucas2025swssbexact,
  title = {Exactly solvable dissipative dynamics and one-form strong-to-weak spontaneous symmetry breaking in interacting two-dimensional spin systems},
  author = {S\'a, Lucas and B\'eri, Benjamin},
  journal = {Phys. Rev. B},
  volume = {112},
  issue = {14},
  pages = {144311},
  numpages = {23},
  year = {2025},
  month = {Oct},
  publisher = {American Physical Society},
  doi = {10.1103/1lzb-vp1w},
  url = {https://link.aps.org/doi/10.1103/1lzb-vp1w}
}

@misc{li2026generalizedsymmetryprotectedtopologicalphases,
      title={Generalized symmetry-protected topological phases in mixed states from gauging dualities}, 
      author={Linhao Li and Zhen Bi and Weiguang Cao},
      year={2026},
      eprint={2603.17282},
      archivePrefix={arXiv},
      primaryClass={cond-mat.str-el},
      url={https://arxiv.org/abs/2603.17282}, 
}

@article{Xu2025averageexacta,
  title = {Average-exact mixed anomalies and compatible phases},
  author = {Xu, Yichen and Jian, Chao-Ming},
  journal = {Phys. Rev. B},
  volume = {111},
  issue = {12},
  pages = {125128},
  numpages = {20},
  year = {2025},
  month = {Mar},
  publisher = {American Physical Society},
  doi = {10.1103/PhysRevB.111.125128},
  url = {https://link.aps.org/doi/10.1103/PhysRevB.111.125128}
}

@article{feng2025random,
  title = {Hardness of Observing Strong-to-Weak Symmetry Breaking},
  author = {Feng, Xiaozhou and Cheng, Zihan and Ippoliti, Matteo},
  journal = {Phys. Rev. Lett.},
  volume = {135},
  issue = {20},
  pages = {200402},
  numpages = {7},
  year = {2025},
  month = {Nov},
  publisher = {American Physical Society},
  doi = {10.1103/1xzd-g9s5},
  url = {https://link.aps.org/doi/10.1103/1xzd-g9s5}
}

@misc{hauser2026strongtoweaksymmetrybreakingopen,
      title={Strong-to-Weak Symmetry Breaking in Open Quantum Systems: From Discrete Particles to Continuum Hydrodynamics}, 
      author={Jacob Hauser and Kaixiang Su and Hyunsoo Ha and Jerome Lloyd and Thomas G. Kiely and Romain Vasseur and Sarang Gopalakrishnan and Cenke Xu and Matthew P. A. Fisher},
      year={2026},
      eprint={2602.16045},
      archivePrefix={arXiv},
      primaryClass={quant-ph},
      url={https://arxiv.org/abs/2602.16045}, 
}

@misc{shu2026universaldynamicalscalingstrongtoweak,
      title={Universal Dynamical Scaling of Strong-to-Weak Spontaneous Symmetry Breaking in Open Quantum Systems}, 
      author={Chang Shu and Kai Zhang and Kai Sun},
      year={2026},
      eprint={2603.06363},
      archivePrefix={arXiv},
      primaryClass={cond-mat.mes-hall},
      url={https://arxiv.org/abs/2603.06363}, 
}

@misc{chen2025zippingmanybodyquantumstates,
      title={Zipping many-body quantum states: a scalable approach to diagonal entropy}, 
      author={Yu-Hsueh Chen and Tarun Grover},
      year={2025},
      eprint={2502.18898},
      archivePrefix={arXiv},
      primaryClass={quant-ph},
      url={https://arxiv.org/abs/2502.18898}, 
}

@Article{Hsin2024,
author={Hsin, Po-Shen
and Luo, Zhu-Xi
and Sun, Hao-Yu},
title={Anomalies of average symmetries: entanglement and open quantum systems},
journal={Journal of High Energy Physics},
year={2024},
month={Oct},
day={18},
volume={2024},
number={10},
pages={134},
issn={1029-8479},
doi={10.1007/JHEP10(2024)134},
url={https://doi.org/10.1007/JHEP10(2024)134}
}

@article{wang2025anomalymixed,
  title = {Anomaly in Open Quantum Systems and its Implications on Mixed-State Quantum Phases},
  author = {Wang, Zijian and Li, Linhao},
  journal = {PRX Quantum},
  volume = {6},
  issue = {1},
  pages = {010347},
  numpages = {27},
  year = {2025},
  month = {Mar},
  publisher = {American Physical Society},
  doi = {10.1103/PRXQuantum.6.010347},
  url = {https://link.aps.org/doi/10.1103/PRXQuantum.6.010347}
}

@article{hasting2011finiteTtopo,
  title = {Topological Order at Nonzero Temperature},
  author = {Hastings, Matthew B.},
  journal = {Phys. Rev. Lett.},
  volume = {107},
  issue = {21},
  pages = {210501},
  numpages = {5},
  year = {2011},
  month = {Nov},
  publisher = {American Physical Society},
  doi = {10.1103/PhysRevLett.107.210501},
  url = {https://link.aps.org/doi/10.1103/PhysRevLett.107.210501}
}

@misc{sang2025mixedstatephaseslocalreversibility,
      title={Mixed-state phases from local reversibility}, 
      author={Shengqi Sang and Leonardo A. Lessa and Roger S. K. Mong and Tarun Grover and Chong Wang and Timothy H. Hsieh},
      year={2025},
      eprint={2507.02292},
      archivePrefix={arXiv},
      primaryClass={quant-ph},
      url={https://arxiv.org/abs/2507.02292}, 
}

@article{Sang2024mixedRG,
  title = {Mixed-State Quantum Phases: Renormalization and Quantum Error Correction},
  author = {Sang, Shengqi and Zou, Yijian and Hsieh, Timothy H.},
  journal = {Phys. Rev. X},
  volume = {14},
  issue = {3},
  pages = {031044},
  numpages = {24},
  year = {2024},
  month = {Sep},
  publisher = {American Physical Society},
  doi = {10.1103/PhysRevX.14.031044},
  url = {https://link.aps.org/doi/10.1103/PhysRevX.14.031044}
}

@Article{ElShowk2014cb_ising,
author={El-Showk, Sheer
and Paulos, Miguel F.
and Poland, David
and Rychkov, Slava
and Simmons-Duffin, David
and Vichi, Alessandro},
title={Solving the 3d Ising Model with the Conformal Bootstrap II. $c$-Minimization and Precise Critical Exponents},
journal={Journal of Statistical Physics},
year={2014},
month={Dec},
day={01},
volume={157},
number={4},
pages={869-914},
issn={1572-9613},
doi={10.1007/s10955-014-1042-7},
url={https://doi.org/10.1007/s10955-014-1042-7}
}

@article{zhuw2023fuzzysphere,
  title = {Uncovering Conformal Symmetry in the 3D Ising Transition: State-Operator Correspondence from a Quantum Fuzzy Sphere Regularization},
  author = {Zhu, Wei and Han, Chao and Huffman, Emilie and Hofmann, Johannes S. and He, Yin-Chen},
  journal = {Phys. Rev. X},
  volume = {13},
  issue = {2},
  pages = {021009},
  numpages = {16},
  year = {2023},
  month = {Apr},
  publisher = {American Physical Society},
  doi = {10.1103/PhysRevX.13.021009},
  url = {https://link.aps.org/doi/10.1103/PhysRevX.13.021009}
}

@article{binder1981renorm,
  title = {Critical Properties from Monte Carlo Coarse Graining and Renormalization},
  author = {Binder, K.},
  journal = {Phys. Rev. Lett.},
  volume = {47},
  issue = {9},
  pages = {693--696},
  numpages = {0},
  year = {1981},
  month = {Aug},
  publisher = {American Physical Society},
  doi = {10.1103/PhysRevLett.47.693},
  url = {https://link.aps.org/doi/10.1103/PhysRevLett.47.693}
}

@Inbook{Melko2013sse,
author="Melko, Roger G.",
editor="Avella, Adolfo
and Mancini, Ferdinando",
title="Stochastic Series Expansion Quantum Monte Carlo",
bookTitle="Strongly Correlated Systems: Numerical Methods",
year="2013",
publisher="Springer Berlin Heidelberg",
address="Berlin, Heidelberg",
pages="185--206",
isbn="978-3-642-35106-8",
doi="10.1007/978-3-642-35106-8_7",
url="https://doi.org/10.1007/978-3-642-35106-8_7"
}

@article{McGreevy2023symmetryreview,
   author = "McGreevy, John",
   title = "Generalized Symmetries in Condensed Matter", 
   journal= "Annual Review of Condensed Matter Physics",
   year = "2023",
   volume = "14",
   number = "Volume 14, 2023",
   pages = "57-82",
   doi = "https://doi.org/10.1146/annurev-conmatphys-040721-021029",
   url = "https://www.annualreviews.org/content/journals/10.1146/annurev-conmatphys-040721-021029",
   publisher = "Annual Reviews",
   issn = "1947-5462",
   type = "Journal Article",
  }

@article{garratt2023,
  title = {Measurements Conspire Nonlocally to Restructure Critical Quantum States},
  author = {Garratt, Samuel J. and Weinstein, Zack and Altman, Ehud},
  journal = {Phys. Rev. X},
  volume = {13},
  issue = {2},
  pages = {021026},
  numpages = {24},
  year = {2023},
  month = {May},
  publisher = {American Physical Society},
  doi = {10.1103/PhysRevX.13.021026},
  url = {https://link.aps.org/doi/10.1103/PhysRevX.13.021026}
}

@article{DELFINO2004521,
title = {Universal ratios along a line of critical points. The Ashkin–Teller model},
journal = {Nuclear Physics B},
volume = {682},
number = {3},
pages = {521-550},
year = {2004},
issn = {0550-3213},
doi = {https://doi.org/10.1016/j.nuclphysb.2004.01.007},
url = {https://www.sciencedirect.com/science/article/pii/S0550321304000227},
author = {Gesualdo Delfino and Paolo Grinza}
}

@article{Komargodski:2016auf,
    author = "Komargodski, Zohar and Simmons-Duffin, David",
    title = "{The Random-Bond Ising Model in 2.01 and 3 Dimensions}",
    eprint = "1603.04444",
    archivePrefix = "arXiv",
    primaryClass = "hep-th",
    doi = "10.1088/1751-8121/aa6087",
    journal = "J. Phys. A",
    volume = "50",
    number = "15",
    pages = "154001",
    year={2016} }

@article{PhysRevB.109.195420,
  title = {Two-dimensional symmetry-protected topological phases and transitions in open quantum systems},
  author = {Guo, Yuxuan and Ashida, Yuto},
  journal = {Phys. Rev. B},
  volume = {109},
  issue = {19},
  pages = {195420},
  numpages = {10},
  year = {2024},
  month = {May},
  publisher = {American Physical Society},
  doi = {10.1103/PhysRevB.109.195420},
  url = {https://link.aps.org/doi/10.1103/PhysRevB.109.195420}
}

@article{PhysRevB.111.115123,
  title = {Mixed-state phase transitions in spin-Holstein models},
  author = {Min, Brett and Zhang, Yuxuan and Guo, Yuxuan and Segal, Dvira and Ashida, Yuto},
  journal = {Phys. Rev. B},
  volume = {111},
  issue = {11},
  pages = {115123},
  numpages = {17},
  year = {2025},
  month = {Mar},
  publisher = {American Physical Society},
  doi = {10.1103/PhysRevB.111.115123},
  url = {https://link.aps.org/doi/10.1103/PhysRevB.111.115123}
}

@article{Ogunnaike:2023qyh,
    author = "Ogunnaike, Olumakinde and Feldmeier, Johannes and Lee, Jong Yeon",
    title = "{Unifying Emergent Hydrodynamics and Lindbladian Low-Energy Spectra across Symmetries, Constraints, and Long-Range Interactions}",
    eprint = "2304.13028",
    archivePrefix = "arXiv",
    primaryClass = "cond-mat.str-el",
    doi = "10.1103/PhysRevLett.131.220403",
    journal = "Phys. Rev. Lett.",
    volume = "131",
    number = "22",
    pages = "220403",
    year = "2023"
}

@article{Lee:2022hog,
    author = "Lee, Jong Yeon and You, Yi-Zhuang and Xu, Cenke",
    title = "{Symmetry protected topological phases under decoherence}",
    eprint = "2210.16323",
    archivePrefix = "arXiv",
    primaryClass = "cond-mat.str-el",
    doi = "10.22331/q-2025-01-23-1607",
    journal = "Quantum",
    volume = "9",
    pages = "1607",
    year = "2025"
}

@article{PhysRevB.24.5229,
  title = {Hamiltonian studies of the $d=2$ Ashkin-Teller model},
  author = {Kohmoto, Mahito and den Nijs, Marcel and Kadanoff, Leo P.},
  journal = {Phys. Rev. B},
  volume = {24},
  issue = {9},
  pages = {5229--5241},
  numpages = {0},
  year = {1981},
  month = {Nov},
  publisher = {American Physical Society},
  doi = {10.1103/PhysRevB.24.5229},
  url = {https://link.aps.org/doi/10.1103/PhysRevB.24.5229}
}

@article{Bao:2023zry,
    author = "Bao, Yimu and Fan, Ruihua and Vishwanath, Ashvin and Altman, Ehud",
    title = "{Mixed-state topological order and the errorfield double formulation of decoherence-induced transitions}",
    eprint = "2301.05687",
    journal = "arXiv",
    month = "1",
    year = "2023"
}

@article{Wang:2025awb,
    author = "Wang, Zijian and Fan, Ruihua and Wang, Tianle and Garratt, Samuel J. and Altman, Ehud",
    title = "{Fractional quantum Hall states under density decoherence}",
    eprint = "2510.08490",
    journal = "arXiv",
    month = "10",
    year = "2025"
}

@article{melko2010ee,
  title = {Finite-size scaling of mutual information in Monte Carlo simulations: Application to the spin-$\frac{1}{2}$ $XXZ$ model},
  author = {Melko, Roger G. and Kallin, Ann B. and Hastings, Matthew B.},
  journal = {Phys. Rev. B},
  volume = {82},
  issue = {10},
  pages = {100409},
  numpages = {4},
  year = {2010},
  month = {Sep},
  publisher = {American Physical Society},
  doi = {10.1103/PhysRevB.82.100409},
  url = {https://link.aps.org/doi/10.1103/PhysRevB.82.100409}
}

@article{ding2024reweight,
  title = {Reweight-annealing method for evaluating the partition function via quantum Monte Carlo calculations},
  author = {Ding, Yi-Ming and Sun, Jun-Song and Ma, Nvsen and Pan, Gaopei and Cheng, Chen and Yan, Zheng},
  journal = {Phys. Rev. B},
  volume = {110},
  issue = {16},
  pages = {165152},
  numpages = {10},
  year = {2024},
  month = {Oct},
  publisher = {American Physical Society},
  doi = {10.1103/PhysRevB.110.165152},
  url = {https://link.aps.org/doi/10.1103/PhysRevB.110.165152}
}

@Article{Wang2025reweight,
author={Wang, Zhe
and Wang, Zhiyan
and Ding, Yi-Ming
and Mao, Bin-Bin
and Yan, Zheng},
title={Bipartite reweight-annealing algorithm of quantum Monte Carlo to extract large-scale data of entanglement entropy and its derivative},
journal={Nature Communications},
year={2025},
month={Jul},
day={01},
volume={16},
number={1},
pages={5880},
issn={2041-1723},
doi={10.1038/s41467-025-61084-7},
url={https://doi.org/10.1038/s41467-025-61084-7}
}

@article{DEmidio2020jazynski,
  title = {Entanglement Entropy from Nonequilibrium Work},
  author = {D'Emidio, Jonathan},
  journal = {Phys. Rev. Lett.},
  volume = {124},
  issue = {11},
  pages = {110602},
  numpages = {5},
  year = {2020},
  month = {Mar},
  publisher = {American Physical Society},
  doi = {10.1103/PhysRevLett.124.110602},
  url = {https://link.aps.org/doi/10.1103/PhysRevLett.124.110602}
}

@article{tarabunga2025bellqmc,
  title = {Bell Sampling in Quantum Monte Carlo Simulations},
  author = {Tarabunga, Poetri Sonya and Ding, Yi-Ming},
  journal = {Phys. Rev. Lett.},
  volume = {135},
  issue = {20},
  pages = {200403},
  numpages = {8},
  year = {2025},
  month = {Nov},
  publisher = {American Physical Society},
  doi = {10.1103/fq8z-y55j},
  url = {https://link.aps.org/doi/10.1103/fq8z-y55j}
}

@article{Swendsen1987,
  title = {Nonuniversal critical dynamics in Monte Carlo simulations},
  author = {Swendsen, Robert H. and Wang, Jian-Sheng},
  journal = {Phys. Rev. Lett.},
  volume = {58},
  issue = {2},
  pages = {86--88},
  numpages = {0},
  year = {1987},
  month = {Jan},
  publisher = {American Physical Society},
  doi = {10.1103/PhysRevLett.58.86},
  url = {https://link.aps.org/doi/10.1103/PhysRevLett.58.86}
}

@article{Kawashima2004worldline,
  author  = {Naoki Kawashima and Kenji Harada},
  title   = {Recent Developments of World-Line Monte Carlo Methods},
  journal = {Journal of the Physical Society of Japan},
  volume  = {73},
  number  = {6},
  pages   = {1379--1414},
  year    = {2004},
  doi     = {10.1143/JPSJ.73.1379}
}

@article{Beard1996continuosqmc,
  title = {Simulations of Discrete Quantum Systems in Continuous Euclidean Time},
  author = {Beard, B. B. and Wiese, U.-J.},
  journal = {Phys. Rev. Lett.},
  volume = {77},
  issue = {25},
  pages = {5130--5133},
  numpages = {0},
  year = {1996},
  month = {Dec},
  publisher = {American Physical Society},
  doi = {10.1103/PhysRevLett.77.5130},
  url = {https://link.aps.org/doi/10.1103/PhysRevLett.77.5130}
}

@article{sandvik1999sse,
  title = {Stochastic series expansion method with operator-loop update},
  author = {Sandvik, Anders W.},
  journal = {Phys. Rev. B},
  volume = {59},
  issue = {22},
  pages = {R14157--R14160},
  numpages = {0},
  year = {1999},
  month = {Jun},
  publisher = {American Physical Society},
  doi = {10.1103/PhysRevB.59.R14157},
  url = {https://link.aps.org/doi/10.1103/PhysRevB.59.R14157}
}

@article{singh2011mutualinformation,
  title = {Finite-Temperature Critical Behavior of Mutual Information},
  author = {Singh, Rajiv R. P. and Hastings, Matthew B. and Kallin, Ann B. and Melko, Roger G.},
  journal = {Phys. Rev. Lett.},
  volume = {106},
  issue = {13},
  pages = {135701},
  numpages = {4},
  year = {2011},
  month = {Mar},
  publisher = {American Physical Society},
  doi = {10.1103/PhysRevLett.106.135701},
  url = {https://link.aps.org/doi/10.1103/PhysRevLett.106.135701}
}

@article{DEmidio2024dqcp,
  title = {Entanglement Entropy and Deconfined Criticality: Emergent SO(5) Symmetry and Proper Lattice Bipartition},
  author = {D'Emidio, Jonathan and Sandvik, Anders W.},
  journal = {Phys. Rev. Lett.},
  volume = {133},
  issue = {16},
  pages = {166702},
  numpages = {7},
  year = {2024},
  month = {Oct},
  publisher = {American Physical Society},
  doi = {10.1103/PhysRevLett.133.166702},
  url = {https://link.aps.org/doi/10.1103/PhysRevLett.133.166702}
}

@article{ding2025negativity,
  title = {Tracking the variation of entanglement R\'enyi negativity: A quantum Monte Carlo study},
  author = {Ding, Yi-Ming and Tang, Yin and Wang, Zhe and Wang, Zhiyan and Mao, Bin-Bin and Yan, Zheng},
  journal = {Phys. Rev. B},
  volume = {111},
  issue = {24},
  pages = {L241108},
  numpages = {7},
  year = {2025},
  month = {Jun},
  publisher = {American Physical Society},
  doi = {10.1103/PhysRevB.111.L241108},
  url = {https://link.aps.org/doi/10.1103/PhysRevB.111.L241108}
}

@article{supplemental,
  journal = {See Supplemental Material for the derivation of $\sigma_1$ and $\sigma_2$, details of the QMC algorithm, and the field-theoretical derivation of the 2D defect action, which includes Refs.~\cite{Beard1996continuosqmc,Kawashima2004worldline,sandvik1999sse,Melko2013sse,sandvik2003sseising,Swendsen1987,Aoun:2024lmd}}
}

@article{Aoun:2024lmd,
    author = "Aoun, Yacine and Dober, Moritz and Glazman, Alexander",
    title = "{Phase Diagram of the Ashkin{\textendash}Teller Model}",
    doi = "10.1007/s00220-023-04925-0",
    journal = "Commun. Math. Phys.",
    volume = "405",
    number = "2",
    pages = "37",
    year = "2024"
}

@misc{wang2026swssb-experiment,
      title={Observation of Strong-to-Weak Spontaneous Symmetry Breaking in a Dephased Fermi Gas}, 
      author={Si Wang and Thomas G. Kiely and Dorothee Tell and Johannes Obermeyer and Marnix Barendregt and Petar Bojović and Philipp M. Preiss and Abhijat Sarma and Titus Franz and Matthew P. A. Fisher and Cenke Xu and Immanuel Bloch},
      year={2026},
      eprint={2604.16137},
      archivePrefix={arXiv},
      primaryClass={cond-mat.quant-gas},
      url={https://arxiv.org/abs/2604.16137}, 
}

@Article{Simon2011tfim-exp,
author={Simon, Jonathan
and Bakr, Waseem S.
and Ma, Ruichao
and Tai, M. Eric
and Preiss, Philipp M.
and Greiner, Markus},
title={Quantum simulation of antiferromagnetic spin chains in an optical lattice},
journal={Nature},
year={2011},
month={Apr},
day={01},
volume={472},
number={7343},
pages={307-312},
issn={1476-4687},
doi={10.1038/nature09994},
url={https://doi.org/10.1038/nature09994}
}

@Article{Labuhn2016tfim-exp,
author={Labuhn, Henning
and Barredo, Daniel
and Ravets, Sylvain
and de L{\'e}s{\'e}leuc, Sylvain
and Macr{\`i}, Tommaso
and Lahaye, Thierry
and Browaeys, Antoine},
title={Tunable two-dimensional arrays of single Rydberg atoms for realizing quantum Ising models},
journal={Nature},
year={2016},
month={Jun},
day={01},
volume={534},
number={7609},
pages={667-670},
issn={1476-4687},
doi={10.1038/nature18274},
url={https://doi.org/10.1038/nature18274}
}

@article{Ware2021pauli-exp,
  title = {Experimental Pauli-frame randomization on a superconducting qubit},
  author = {Ware, Matthew and Ribeill, Guilhem and Rist\`e, Diego and Ryan, Colm A. and Johnson, Blake and da Silva, Marcus P.},
  journal = {Phys. Rev. A},
  volume = {103},
  issue = {4},
  pages = {042604},
  numpages = {9},
  year = {2021},
  month = {Apr},
  publisher = {American Physical Society},
  doi = {10.1103/PhysRevA.103.042604},
  url = {https://link.aps.org/doi/10.1103/PhysRevA.103.042604}
}

@article{Lu2017pauli-exp,
  title = {Experimental quantum channel simulation},
  author = {Lu, He and Liu, Chang and Wang, Dong-Sheng and Chen, Luo-Kan and Li, Zheng-Da and Yao, Xing-Can and Li, Li and Liu, Nai-Le and Peng, Cheng-Zhi and Sanders, Barry C. and Chen, Yu-Ao and Pan, Jian-Wei},
  journal = {Phys. Rev. A},
  volume = {95},
  issue = {4},
  pages = {042310},
  numpages = {8},
  year = {2017},
  month = {Apr},
  publisher = {American Physical Society},
  doi = {10.1103/PhysRevA.95.042310},
  url = {https://link.aps.org/doi/10.1103/PhysRevA.95.042310}
}

@article{Chiuri2011pauli-exp,
  title = {Experimental Realization of Optimal Noise Estimation for a General Pauli Channel},
  author = {Chiuri, A. and Rosati, V. and Vallone, G. and P\'adua, S. and Imai, H. and Giacomini, S. and Macchiavello, C. and Mataloni, P.},
  journal = {Phys. Rev. Lett.},
  volume = {107},
  issue = {25},
  pages = {253602},
  numpages = {5},
  year = {2011},
  month = {Dec},
  publisher = {American Physical Society},
  doi = {10.1103/PhysRevLett.107.253602},
  url = {https://link.aps.org/doi/10.1103/PhysRevLett.107.253602}
}

@article{Daley2012ee-replica-exp,
  title = {Measuring Entanglement Growth in Quench Dynamics of Bosons in an Optical Lattice},
  author = {Daley, A. J. and Pichler, H. and Schachenmayer, J. and Zoller, P.},
  journal = {Phys. Rev. Lett.},
  volume = {109},
  issue = {2},
  pages = {020505},
  numpages = {5},
  year = {2012},
  month = {Jul},
  publisher = {American Physical Society},
  doi = {10.1103/PhysRevLett.109.020505},
  url = {https://link.aps.org/doi/10.1103/PhysRevLett.109.020505}
}

@Article{Islam2015gas-micro-ee,
author={Islam, Rajibul
and Ma, Ruichao
and Preiss, Philipp M.
and Eric Tai, M.
and Lukin, Alexander
and Rispoli, Matthew
and Greiner, Markus},
title={Measuring entanglement entropy in a quantum many-body system},
journal={Nature},
year={2015},
month={Dec},
day={01},
volume={528},
number={7580},
pages={77-83},
issn={1476-4687},
doi={10.1038/nature15750},
url={https://doi.org/10.1038/nature15750}
}

@article{Tiff2019random,
author = {Tiff Brydges  and Andreas Elben  and Petar Jurcevic  and Benoît Vermersch  and Christine Maier  and Ben P. Lanyon  and Peter Zoller  and Rainer Blatt  and Christian F. Roos },
title = {Probing Rényi entanglement entropy via randomized measurements},
journal = {Science},
volume = {364},
number = {6437},
pages = {260-263},
year = {2019},
doi = {10.1126/science.aau4963},
URL = {https://www.science.org/doi/abs/10.1126/science.aau4963},
eprint = {https://www.science.org/doi/pdf/10.1126/science.aau4963}}

\section{End Matter}

\subsection{Weak symmetry and the Rényi-2 linear order correlator in the doubled space}
To illustrate that $C^{(1)}$ may not be a faithful diagnostic of weak symmetry breaking in the mixed state, we construct a diagonal density matrix in the local $Z$ basis:
\begin{equation}
    \begin{aligned}
        \rho &= \frac{1}{4}\Bigl( \ket{0}^{\otimes N}\bra{0}^{\otimes N} + \ket{1}^{\otimes N}\bra{1}^{\otimes N}\Bigr) \\
        &\quad + \frac{1}{4M}\sum_{\alpha=1}^{M}
        \Bigl(
        \ket{\psi_{\alpha}}\bra{\psi_{\alpha}} + 
        X\ket{\psi_{\alpha}}\bra{\psi_{\alpha}}X
        \Bigr),
    \end{aligned}
\end{equation}
where $X:=\prod_i X_i$, and $\{\ket{\psi_{\alpha}}\}$ satisfy $Z_i Z_j \ket{\psi_{\alpha}} = - \ket{\psi_{\alpha}}$.
By construction, $\rho$ is normalized and has a global weak $\mathbb{Z}_2$ symmetry.

A direct calculation gives $C^{(0)}=0$ but $C^{(1)}=(M-1)/(M+1)$. 
Thus, for $M>1$, $C^{(1)}\neq 0$ despite $C^{(0)}=0$, showing that $C^{(1)}$, while natural as a linear order parameter at the Choi-state level, does not provide a faithful diagnostic of weak symmetry breaking in mixed states.

\subsection{Stability of the conventional ferromagnetic order under decoherence}
Here we clarify why the ground-state ordered region $J>J_c$ remains conventionally ferromagnetic under the decoherence channel~\eqref{eq:channel}. For the linear correlator
$C_{xy}^{(0)}(\rho)=\Tr(\rho Z_xZ_y)$, the bond operator $Z_iZ_j$ in the channel~\eqref{eq:channel} commutes with $Z_xZ_y$. Therefore, each local channel leaves the linear correlator unchanged:
\begin{equation}
    C_{xy}^{(0)}\!\left(\mathcal{E}_{\langle ij\rangle}[\rho_0]\right)
    =
    C_{xy}^{(0)}(\rho_0).
\end{equation}
Applying this result successively to all other local channels gives
\begin{equation}
    C_{xy}^{(0)}(\mathcal{E}[\rho_0])
    =
    C_{xy}^{(0)}(\rho_0),
    \qquad
    \mathcal{E}=\prod_{\langle ij\rangle}\mathcal{E}_{\langle ij\rangle}.
\end{equation}
Hence, if the original TFIM ground state is already ordered when
$J>J_c$, then
\begin{equation}
    \lim_{|x-y|\to\infty} C_{xy}^{(0)}(\mathcal{E}[\rho_0])
    =
    \lim_{|x-y|\to\infty} C_{xy}^{(0)}(\rho_0)
    \neq 0 .
\end{equation}
Thus, the decoherence channel cannot restore the symmetry when $J>J_c$, since the linear order parameter $C^{(0)}$ is exactly unchanged.

\begin{figure}[htbp]
        \centering
        \includegraphics[width=\linewidth]{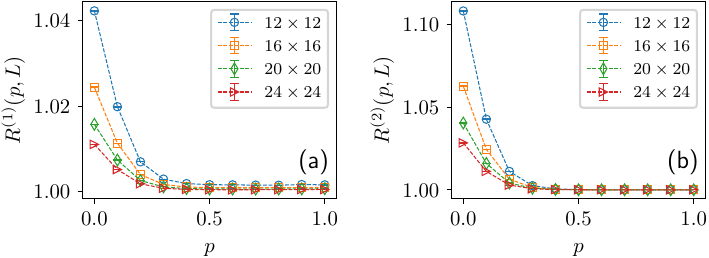}
        \caption{
            Fixing $J=0.4$, scanning $p$, the corresponding binder ratios of the correlators $C^{(1)}$ in (a) and $C^{(2)}$ in (b), respectively.
        }
    \label{fig:added_ordered_phase}
\end{figure}

We further confirm this conclusion numerically by fixing $J=0.4>J_c$ and scanning the decoherence strength $p$ from $0$ to $1$. As shown in Fig.~\ref{fig:added_ordered_phase}, the Binder ratios associated with both $C^{(1)}$ and $C^{(2)}$ rapidly approach the ordered-region value $1$ with increasing system size, and no crossing point is observed. This provides additional evidence that no decoherence-driven phase transition occurs in the conventionally ordered region.

\subsection{Phase transition between R2-SSB and R2-SWSSB}
\begin{figure}[ht!]
    \centering
    \includegraphics[width=\linewidth]{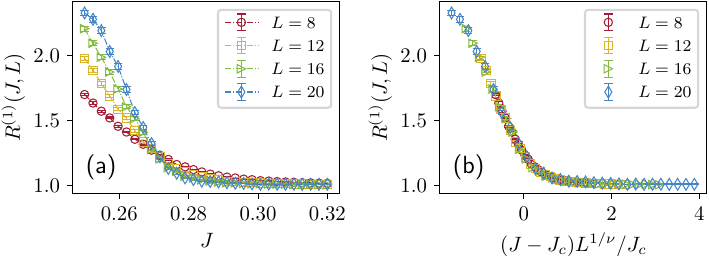}
    \caption{
    An example for the blue critical boundary in Fig.~\ref{fig:phase_diagram}. 
    At fixed $p=0.6$, we tune $J$ and evaluate the Binder ratio $R^{(1)}$. 
    (a) Crossings of $R^{(1)}(J,L)$ for different system sizes locate the critical point at $J_c \approx 0.27$. 
    (b) Finite-size scaling yields an excellent data collapse with $\nu \approx 0.98$, close to the expected Ising value $\nu=1$.
    }
    \label{fig:blue_example}
\end{figure}

In this subsection, we show a representative example of the numerical results on blue critical boundary in Fig.~\ref{fig:phase_diagram}, associated with $C^{(1)}$. We fix $p=0.6$ and tune $J$. 
Fig.~\ref{fig:blue_example} shows crossings of the Binder ratio $R^{(1)}$ for different system sizes, from which we extract the critical point $J_c \approx 0.27$ and the correlation-length exponent $\nu \approx 0.98\approx 1$, consistent with the 2D Ising universality class. 
This provides further numerical support for the Ising nature of the blue phase boundary discussed in the main text.

\subsection{Decoherence at Ising criticality}

\begin{figure}[ht!]
    \centering
    \includegraphics[width=\linewidth]{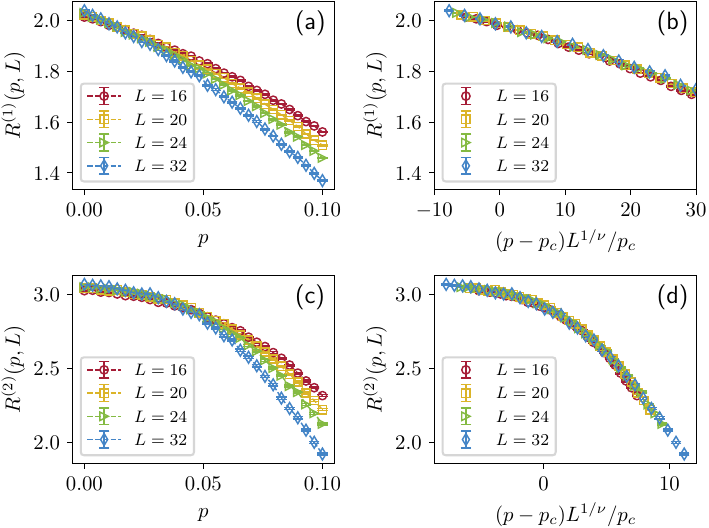}
    \caption{
    Binder ratios $R^{(\alpha)}$ as functions of system size $L$ and tuning parameters $J$ and $p$.
    Results for $\alpha=1$ are shown in (a, b), and for $\alpha=2$ in (c, d).
    We fix $J = 0.328474$.
    (a, c) show $R^{(\alpha)}$ versus the decoherence rate $p$ for different $L$, exhibiting crossings that locate the critical points, while (b, d) show the corresponding data collapse using $\nu= 1.70$.
    The crossing points are $p_c \approx 0.01$ in (a) and $p_c \approx 0.04$ in (c).
    }
    \label{fig:pertb_3d_ising}
\end{figure}

We now analyze the regime in which the bulk is tuned to criticality, such that the defect is coupled to a 3D Ising CFT. In this case, the effective description in terms of a 2D short-range Ashkin-Teller model breaks down due to the emergence of long-range interactions.
We therefore analyze this regime using conformal perturbation theory \cite{Komargodski:2016auf}.

The leading relevant perturbation on the defect is given by
\begin{equation}
    -\tilde{m}\int d^2x\, dz\, \delta(z)\, (\varepsilon^a+\varepsilon^b),
\end{equation}
where $\varepsilon$ denotes the energy-density operator. 
Given the scaling dimension $\Delta_\varepsilon \approx 1.41 < 2$ for the 3D Ising universality class~\cite{ElShowk2014cb_ising,zhuw2023fuzzysphere}, this perturbation is relevant.
It drives the theory into the R2-SSB  phase, breaking $\mathbb{Z}^a_2 \times \mathbb{Z}^b_2$ down to the trivial group. Consequently, the phase transition occurs exactly at $p=0$, characterized by the critical exponent $\nu = 1/(2-\Delta_\varepsilon)\approx 1.70$.

To verify this numerically, we fix $J=0.328474\approx J_c$~\cite{Henk2022qmc_ising} and vary $p$, measuring both the double-space Rényi-2 linear Binder ratio $R^{(1)}$ and the Rényi-2 Binder ratio $R^{(2)}$. As shown in Fig.~\ref{fig:pertb_3d_ising}(a, c), finite-size effects near criticality induce a systematic drift of the crossing points at fixed $J=J_c$, yielding $p_c\approx 0.01$ for $R^{(1)}$ and $p_c\approx 0.04$ for $R^{(2)}$. Nevertheless, we find that with increasing system size the crossings move consistently toward the expected $p_c=0$. 

The finite-size scaling analysis using the theoretical exponent $\nu = 1.70$ yields a clear and good data collapse [Fig.~\ref{fig:pertb_3d_ising}(b), (d)], capturing the overall scaling behavior across the accessible system sizes. This provides strong evidence that the numerical results are consistent with the field-theoretical prediction. Independent fits give $\nu \approx 2.23$ for $R^{(1)}$ and $\nu \approx 1.87$ for $R^{(2)}$, with the latter already close to the expected value. The larger deviation in $R^{(1)}$ can be attributed to stronger finite-size effects and statistical fluctuations. Overall, these results support the expected universality class, with remaining discrepancies attributable to finite-size corrections for system sizes up to $32 \times 32$.

\subsection{Continuously varying critical exponents}
\begin{figure}[htbp]
    \centering
    \includegraphics[width=\linewidth]{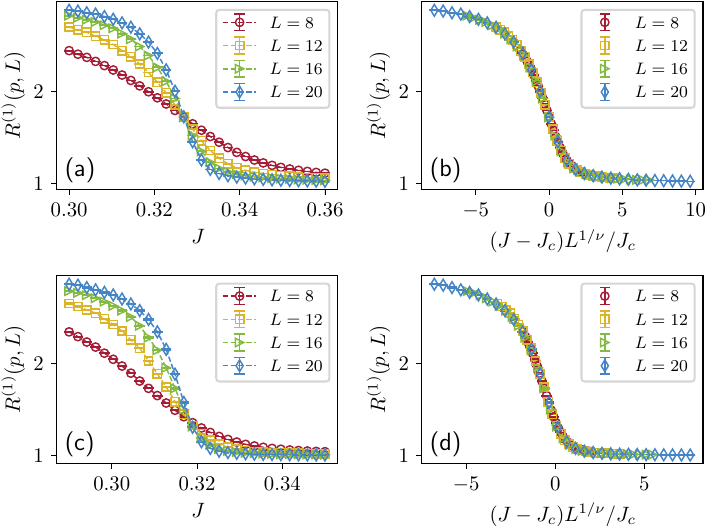}
    \caption{
    Representative examples along the overlapping critical boundary separating the strongly symmetric and R2-SSB phases in Fig.~\ref{fig:phase_diagram}. 
    At fixed $p=0.1$ [(a), (b)] and $p=0.2$ [(c), (d)], we tune $J$ and evaluate the Binder ratio $R^{(1)}$. 
    (a), (c) Crossings of $R^{(1)}(J,L)$ for different system sizes locate the critical points at $J_c \approx 0.327$ ($p=0.1$) and $J_c \approx 0.318$ ($p=0.2$). 
    (b), (d) Finite-size scaling yields good data collapse with $\nu \approx 0.658$ ($p=0.1$) and $\nu \approx 0.691$ ($p=0.2$), demonstrating continuously varying critical exponents along this phase boundary.
    }
    \label{fig:vary_exponent}
\end{figure}
In this subsection, we present representative numerical results for the overlapping critical boundary in Fig.~\ref{fig:phase_diagram}, separating the strongly symmetric and R2-SSB phases, where the critical behavior is expected to exhibit continuously varying exponents. 

We fix $p=0.1$ and $p=0.2$, and tune $J$ to probe the phase transitions, as shown in Fig.~\ref{fig:vary_exponent}.
For $p=0.1$ and $p=0.2$, crossings of the Binder ratio $R^{(1)}(J,L)$ locate the transition points at $J_c \approx 0.327$ and $J_c \approx 0.318$, respectively. 
Performing finite-size scaling around these points yields correlation-length exponents $\nu \approx 0.658$ and $\nu \approx 0.691$, demonstrating a clear dependence of the critical exponent on $p$ and providing direct numerical evidence for continuously varying criticality along this phase boundary.

\clearpage
\onecolumngrid
\setcounter{page}{1}

\begin{center}
\textbf{\large Supplemental Material for ``Strong-to-Weak Spontaneous Symmetry Breaking \\in a $(2+1)$D Transverse-Field Ising Model under Decoherence''}\\[2ex]
Yi-Ming Ding, Yuxuan Guo, Zhen Bi, and Zheng Yan
\end{center}

\setcounter{section}{0}
\setcounter{equation}{0}
\setcounter{figure}{0}

\renewcommand{\thesection}{S\arabic{section}}
\renewcommand{\theequation}{S\arabic{equation}}
\renewcommand{\thefigure}{S\arabic{figure}}

\section{Strong-to-weak spontaneous symmetry breaking}
\label{app:swssb}

In this section, we briefly review the notions of weak and strong symmetry for mixed states and quantum channels.  
Building on these definitions, we introduce the diagnostic
observables used in the main text in the next section.

The distinction is essential for mixed states: a density matrix may be invariant as an operator even when its decomposition involves components from different symmetry sectors.  This gives rise to the possibility of strong-to-weak spontaneous symmetry breaking (SWSSB), where the strong symmetry is spontaneously broken while the weak symmetry remains intact.

Specifically, let $G$ be an internal symmetry group represented on the Hilbert space $\mathcal{H}$ by unitary operators $U_g$, with $g\in G$. For convenience, we define the following superoperators:
\begin{equation}
    \mathcal{L}_g(\rho):=U_g\rho,\qquad
    \mathcal{R}_g(\rho):=\rho U_g^\dagger,\qquad
    \mathcal{U}_g(\rho):=U_g\rho U_g^\dagger .
    \label{eq:lr_adjoint_actions}
\end{equation}
Here $\mathcal{L}_g$ and $\mathcal{R}_g$ are the left and right symmetry
actions, while $\mathcal{U}_g=\mathcal{L}_g\circ\mathcal{R}_g$ is the usual
adjoint action.

Two mixed-state symmetries are defined as follows:
\begin{itemize}
    \item A mixed state $\rho$ has a \emph{weak} $G$ symmetry if it is invariant under the adjoint action,
    \begin{equation}
      \mathcal{U}_g(\rho)=  U_g \rho U_g^\dagger=\rho,
        \qquad \forall g\in G .
        \label{eq:weak_state_symmetry}
    \end{equation}
    Equivalently, $[\rho,U_g]=0$ for all $g\in G$;
    \item A mixed state $\rho$ has a \emph{strong} $G$ symmetry if the symmetry acts separately on the left and right of $\rho$.  For the one-dimensional charge sectors relevant to the present work, this condition can be written as
    \begin{equation}
      \mathcal{L}_g(\rho)=  U_g \rho =  e^{\mathrm{i}\phi_g}\rho,
        \qquad
        \mathcal{R}_g(\rho) = \rho U_g^\dagger = e^{-\mathrm{i}\phi_g}\rho,
        \qquad \forall g\in G,
        \label{eq:strong_state_symmetry}
    \end{equation}
    where $e^{\mathrm{i}\phi_g}$ is a one-dimensional character of $G$.  
\end{itemize}

 Note that Eq.~\eqref{eq:strong_state_symmetry} immediately implies Eq.~\eqref{eq:weak_state_symmetry}, thus strong symmetry is a refinement of weak symmetry.  
 Weak symmetry is therefore an operator- or ensemble-level condition: only the
full density matrix is required to be symmetric.
 Physically, a strongly symmetric mixed state is analogous to a fixed-charge ensemble, while a weakly symmetric mixed state may contain an incoherent mixture of different charge sectors as long as the density matrix as a whole is invariant.

The symmetry properties of a quantum channel are the dynamical counterparts of the
state symmetries defined above:
\begin{itemize}
    \item A quantum channel $\mathcal{E}$ has a weak $G$ symmetry if it is covariant under the adjoint action,
    \begin{equation}
        \mathcal{E}\circ \mathcal{U}_g
        =
        \mathcal{U}_ g\circ \mathcal{E},
        \qquad \forall g\in G .
    \end{equation}
     Equivalently,
    \begin{equation}
        \mathcal{E}\!\left(U_g\rho U_g^\dagger\right)
        =
        U_g\mathcal{E}(\rho)U_g^\dagger,
        \qquad \forall \rho,\ \forall g\in G .
    \end{equation}
    \item A quantum channel $\mathcal{E}$ has a strong $G$ symmetry if it commutes with the left and right symmetry actions separately,
    \begin{equation}
         \mathcal{E}\circ\mathcal{L}_g
        =
        \mathcal{L}_g\circ\mathcal{E},
        \qquad
        \mathcal{E}\circ\mathcal{R}_g
        =
        \mathcal{R}_g\circ\mathcal{E},
        \qquad \forall g\in G .
    \end{equation}
    Equivalently,
    \begin{equation}
        \mathcal{E}(U_g\rho)=U_g\mathcal{E}(\rho),
        \qquad
        \mathcal{E}(\rho U_g^\dagger)=\mathcal{E}(\rho)U_g^\dagger,
        \qquad \forall \rho,\ \forall g\in G .
        \label{eq:chnn-strong-sym-def}
    \end{equation}
\end{itemize}

In our setup, the ground state $\rho_0$ of the transverse-field Ising model (TFIM) is strongly $\mathbb{Z}_2$ symmetric under
$X\equiv \prod_i X_i$ in the paramagnetic phase. In the ferromagnetic phase, this
strong symmetry is spontaneously broken to the trivial group $\{e\}$, which is the ordinary (strong-to-trivial) \emph{spontaneous symmetry breaking (SSB)}.
Moreover, one can easily verify that the decoherence channel $\mathcal{E}$ is itself strongly
$\mathbb{Z}_2$ symmetric according to Eq.~\eqref{eq:chnn-strong-sym-def}. Hence, when acting on a strongly symmetric input
state $\rho_0$, the decohered state $\mathcal{E}(\rho_0)$ still formally preserves the
strong $\mathbb{Z}_2$ symmetry.

To illustrate how SWSSB can arise, consider the product-state limit $J=0$ of
the TFIM and the fully decohered limit $p=1$. The initial ground state is
\begin{equation}
    \rho_0
    =
    \ket{+}^{\otimes N}\bra{+}^{\otimes N}
    =
    \frac{1}{2^N}
    \prod_{i=1}^N \left(I_i+X_i\right),
    \qquad
    \ket{+}=\frac{\ket{0}+\ket{1}}{\sqrt{2}},
\end{equation}
where $I_i$ and $X_i$ are the identity and Pauli-$X$ operators on site $i$.
The decoherence channel is
\begin{equation}
    \mathcal{E}
    =
    \prod_{\langle ij\rangle}\mathcal{E}_{\langle ij\rangle},
    \qquad
    \mathcal{E}_{\langle ij\rangle}(\rho)
    =
    \frac{1}{2}
    \left(
        \rho+Z_iZ_j\rho Z_iZ_j
    \right).
\end{equation}
This gives the decohered state
\begin{equation}\label{eq:rho_plus_def}
    \rho_+
    \equiv
    \mathcal{E}(\rho_0)
    =
    \frac{I+X}{2^N},\quad I\equiv\prod_i I_i.
\end{equation}
To expose its structure,
we rewrite Eq.~\eqref{eq:rho_plus_def} in the $Z$ basis. Let
$s\in\{0,1\}^{\otimes N} \equiv \{\uparrow,\downarrow\}^{\otimes N}$ denote a bit string, and let $\bar{s}$ denote its globally
flipped partner, i.e., $\ket{\bar{s}}=X\ket{s}$. Then
\begin{align}
    \rho_+ \propto \sum_s \ket{\mathrm{Cat}_s}\bra{\mathrm{Cat}_s} \label{eq:rho_plus},\quad 
     \ket{\mathrm{Cat}}_s = \frac{\ket{s}+\ket{\bar s}}{\sqrt{2}}.
\end{align}
Thus $\rho_+$ is a uniform convex mixture of cat states, or equivalently the maximally mixed state within the $X=+1$ symmetry sector.

Eq.~\eqref{eq:rho_plus} explains why the formal strong $\mathbb{Z}_2$ symmetry of $\rho_+$ is compatible with the behavior of SWSSB. Each $\ket{\mathrm{Cat}_s}$ is
formally symmetric, just like the Ising cat state for describing a quantum ferromagnet
\begin{equation}
    \ket{\mathrm{Cat}}
    =
    \frac{\ket{0\cdots 0}+\ket{1\cdots 1}}{\sqrt{2}},
\end{equation}
which is invariant under the global strong $\mathbb{Z}_2$ symmetry. In the
thermodynamic limit ($N\to\infty$), however, $\ket{\mathrm{Cat}}$ is unstable to infinitesimal
longitudinal symmetry-breaking fields $h_l$, which select a symmetry-broken
configuration. 
Mathematically, this reflects the noncommutativity
of the limits, $\lim_{h_l\to 0}\lim_{N\to\infty} \neq \lim_{N\to\infty}\lim_{h_l\to 0}$.
Similarly, $\rho_+$ remains formally strongly symmetric at finite
size, but its strong symmetry is spontaneously broken to the weak one in the thermodynamic limit.

We now explain why SWSSB cannot be diagnosed by an ordinary linear correlator. For the SWSSB state $\rho_+$ in Eq.~\eqref{eq:rho_plus}, one finds
\begin{equation}
    C^{(0)}_{ij}
    =
    \Tr(\rho_+ Z_iZ_j)
    =
    \frac{1}{2^N}
    \left[
        \Tr(Z_iZ_j)
        +
        \Tr\!\left(X Z_iZ_j\right)
    \right]
    =
    0, \label{eq:rho_plus_C0}
\end{equation}
using the fact that every nontrivial Pauli string is traceless. 
Therefore, as long as the weak symmetry is preserved, $C^{(0)}_{ij}$ cannot
diagnose whether the strong $\mathbb{Z}_2$ symmetry is broken.

In more general cases, for a symmetry group $G$, let $O_\alpha$ be a local order-parameter multiplet, then it transforms as 
\begin{equation}
U_g^\dagger O_\alpha U_g
=
\sum_\beta R_{\alpha\beta}(g) O_\beta ,
\end{equation}
where $R(g)$ is the corresponding representation matrix.
If the density matrix has weak symmetry, i.e., $U_g\rho U_g^\dagger=\rho$, then the linear correlator gives 
\begin{equation}
C^{(0)}_\alpha=\Tr(\rho O_\alpha) 
=
\Tr(U_g\rho U_g^\dagger O_\alpha)
=
\sum_\beta R_{\alpha\beta}(g)C^{(0)}_\beta ,
\qquad \forall g\in G .
\end{equation}
The vector $C^{(0)}_\alpha$ must be unchanged by every weak symmetry operation. Averaging over all $g\in G$ gives
\begin{equation}
C^{(0)}_\alpha
=
\frac{1}{|G|}\sum_\beta
\sum_{g\in G}R_{\alpha\beta}(g)C^{(0)}_\beta .
\end{equation}
For a nontrivial irreducible representation, the group average of matix element $R_{\alpha\beta}$ vanishes.  Hence a weakly symmetric density matrix has $C^{(0)}_\alpha=0$ for any nontrivial order parameter. 
However, this constraint does not apply to nonlinear quantities, including the
Rényi-2 observables introduced below.

\section{Rényi-$2$ observables}
Consider the Choi--Jamio\l{}kowski isomorphism~\cite{Choi1975,Jamiolkowski1972},
which maps operators on the Hilbert space $\mathcal{H}$ to vectors in the doubled Hilbert space $\mathcal{H}_a\otimes\mathcal{H}_b^*$, where the subscripts $a$ and $b$ label the two copies. For linear bounded operator 
$\rho\in\mathcal{B}(\mathcal{H})$, with matrix elements
$\rho_{ss'}$ in an orthonormal basis $\{\ket{s}\}$,
\begin{equation}
    \rho
    =
    \sum_{s,s'} \rho_{ss'} \ket{s}\!\bra{s'}
    \quad
    \longmapsto
    \quad
    \ket{\rho}\!\rangle
    =
    \sum_{s,s'} \rho_{ss'} \ket{s}_a \ket{s'}_b^* ,
    \label{eq:choi_map}
\end{equation}
where $|{\rho}\rangle\!\rangle$ is the Choi state of $\rho$, which is a pure state in the doubled space $\mathcal{H}_a\otimes\mathcal{H}^*_b$. 

We again use the TFIM as a concrete example. In the doubled Hilbert space, the
symmetry group is 
\begin{equation}
    \mathbb{Z}_2^a \times \mathbb{Z}_2^b \rtimes \mathbb{Z}_2^H,
\end{equation}
where $\mathbb{Z}_2^H$ originates from the Hermiticity of the original density matrix $\rho$ and exchanges the two spaces. This structure leads to the following hierarchy of symmetries for the Choi state $|\rho\rangle\!\rangle$:
\begin{equation}\label{eq:hierarchy}
\mathbb{Z}_2^a \times \mathbb{Z}_2^b
\supset
\mathbb{Z}_2^{\mathrm{diag}}
\supset
\{e\}.
\end{equation}
Explicitly,
\begin{align}
\mathbb{Z}_2^a \times \mathbb{Z}_2^b
&=
\{(e^a,e^b), (e^a,g^b), (g^a,e^b), (g^a,g^b)\}, \\
\mathbb{Z}_2^{\mathrm{diag}}
&=
\{(e^a,e^b), (g^a,g^b)\}
\cong \mathbb{Z}_2,
\end{align}
with $(g^{a/b})^2=e^{a/b}$.

Since $|\rho\rangle\!\rangle$ is a pure state, it can only have a strong symmetry or be symmetry-breaking (trivial symmetry). Moreover, the hierarchy in Eq.~\eqref{eq:hierarchy} allows the Choi state to realize two levels of strong symmetry, which correspond to different symmetry notions for the original density matrix $\rho$:
\begin{align} 
\text{Strong $\mathbb{Z}_2^a\times\mathbb{Z}_2^b$ symmetry of $|\rho\rangle\!\rangle$} \; &\longleftrightarrow \; \text{Strong $\mathbb{Z}_2$ symmetry of $\rho$}, \\ 
\text{Strong $\mathbb{Z}_2^{\mathrm{diag}}$ symmetry of $|\rho\rangle\!\rangle$} \; &\longleftrightarrow \; \text{Weak $\mathbb{Z}_2$ symmetry of $\rho$}, \label{eq:strong-to-weak}\\ 
\text{Complete symmetry-breaking of $|\rho\rangle\!\rangle$} \; &\longleftrightarrow \; 
\text{Complete symmetry-breaking of $\rho$}. \label{eq:strong-to-trivial}
\end{align}

Here Eq.~\eqref{eq:strong-to-weak} corresponds to the \emph{partial} (strong-to-strong) symmetry breaking $\mathbb{Z}_2^a \times \mathbb{Z}_2^b
\longrightarrow
\mathbb{Z}_2^{\mathrm{diag}}$
of the Choi state $|\rho\rangle\!\rangle$. 
For the density matrix $\rho$, it is interpreted as the Rényi-2 SWSSB. 
At the Choi-state level, the natural diagnostic for SWSSB is the Rényi-2 correlator
\begin{equation}
    C^{(2)}
    =
    \lim_{|i-j|\to\infty}
    \frac{
    \langle\!\langle \rho |
    Z_i^a Z_j^a \otimes Z_i^b Z_j^b
    |\rho\rangle\!\rangle
    }{
    \langle\!\langle \rho|\rho\rangle\!\rangle
    }
    =
    \lim_{|i-j|\to\infty}
    \frac{
    \Tr(\rho Z_iZ_j\rho Z_iZ_j)
    }{
    \Tr(\rho^2)
    } .
    \label{eq:renyi2_order_correlator}
\end{equation}
For the illustrative SWSSB state $\rho_+$ in Eq.~\eqref{eq:rho_plus}, one finds
$C^{(2)}=1\neq 0$. Together with $C^{(0)}=0$ in
Eq.~\eqref{eq:rho_plus_C0}, this shows that the strong symmetry is
spontaneously broken to a weak symmetry.

By contrast, Eq.~\eqref{eq:strong-to-trivial} corresponds to ordinary SSB of the pure state $|\rho\rangle\!\rangle$, where even the smaller diagonal subgroup $\mathbb{Z}_2^{\mathrm{diag}}$ is no longer preserved. 
\textbf{At the Choi-state or Rényi-$2$ level}, this complete breaking is diagnosed by
\begin{equation}
C^{(1)}=\lim_{|i-j|\to\infty}\frac{\langle\!\langle \rho| Z^{a/b}_iZ^{a/b}_j |\rho\rangle\!\rangle}{\langle\!\langle \rho|\rho\rangle\!\rangle} 
=\lim_{|i-j|\to\infty}\frac{\Tr(\rho^2 Z_iZ_j)}{\Tr(\rho^2)}
\ne 0.
\end{equation}

On the other hand, if one considers directly the original density matrix $\rho$ without relying on the Choi-state viewpoint, then complete symmetry breaking of $\rho$ should be diagnosed by the conventional linear correlator
\begin{equation}
C^{(0)}=\lim_{|i-j|\to\infty} \Tr(\rho Z_iZ_j)\ne 0,
\end{equation}
This is the faithful diagnostic of ordinary symmetry breaking \textbf{at the mixed-state level}. Therefore, one should not assume that the Choi-state diagnostic $C^{(1)}$, although natural at the Rényi-2 level, is always fully equivalent to the linear mixed-state diagnostic $C^{(0)}$. The counterexample introduced in the End Matter is to illustrate this distinction: $C^{(1)}$ can diagnose symmetry breaking of the Choi state, but it does not by itself provide a faithful diagnostic of complete (weak) symmetry breaking of the original density matrix. 

In practice, both $C^{(0)}$ and $C^{(1)}$ should be taken into consideration.
More importantly, for the setup considered in our work, $C^{(1)}$ is not merely a Choi-state diagnostic. It has a natural physical interpretation, as it diagnoses one of the Ising phase transitions in the effective
Ashkin--Teller theory.



\section{Derivation of Eqs.~(\ref{eq:sigma1}) and (\ref{eq:sigma2})}
The local channel is $\mathcal{E}_{\langle ij\rangle}=\sum_k M_k\rho M_k^{\dagger}$. 
Here, we rewrite $M_k$ defined in Eq.~\eqref{eq:defm_k} as 
\begin{align}
    M_0 &= \sqrt{1-p}\, \mathbbm{1}_i \mathbbm{1}_j , \\
    M_1 &
        = \sqrt{p}\Big(\ket{0_i 0_j}\bra{0_i 0_j} + X_iX_j\ket{0_i 0_j}\bra{0_i 0_j}X_iX_j\Big) 
        , \\
    M_2 &
        = \sqrt{p}\Big(\ket{0_i 1_j}\bra{0_i 1_j} + X_iX_j\ket{0_i 1_j}\bra{0_i 1_j}X_iX_j \Big)
        .
\end{align}
For a computational basis state $\ket{s}=\otimes_k \ket{s_k}$, we decompose it as
\begin{equation}
    \ket{s} \equiv \ket{s_{ij}}\otimes \ket{s_{\overline{ij}}},
\end{equation}
where
\begin{equation}
    \ket{s_{ij}} \equiv \ket{s_i}\otimes \ket{s_j}, 
    \qquad
    \ket{s_{\overline{ij}}}\equiv \otimes_{k\neq i,j}\ket{s_k}.
\end{equation}
Accordingly, the density matrix $\rho_0$ can be expanded as
\begin{equation}\label{eq:rho0_expand}
    \rho_0
    = \sum_{s,s'} \langle s|\rho_0|s'\rangle \ket{s}\bra{s'}
    = \sum_{s,s'} \langle s_{ij},s_{\overline{ij}}|\rho_0|s'_{ij},s'_{\overline{ij}}\rangle
      \ket{s_{ij},s_{\overline{ij}}}\bra{s'_{ij},s'_{\overline{ij}}}.
\end{equation}

We first evaluate the contribution from $M_1$:
\begin{equation}
    \begin{aligned}
        \frac{1}{p} M_1 \rho_0 M_1^\dagger
        =&\; \ket{0_i0_j}\bra{0_i0_j}\rho_0\ket{0_i0_j}\bra{0_i0_j}
        + \ket{0_i0_j}\bra{0_i0_j}\rho_0\ket{1_i1_j}\bra{1_i1_j} \\
        &+ \ket{1_i1_j}\bra{1_i1_j}\rho_0\ket{0_i0_j}\bra{0_i0_j}
        + \ket{1_i1_j}\bra{1_i1_j}\rho_0\ket{1_i1_j}\bra{1_i1_j}.
    \end{aligned}
\end{equation}
Using the decomposition above, this can be rewritten as

\begin{equation}
    \begin{aligned}
       \frac{1}{p} M_1\rho_0 M_1^{\dagger}
        =& \sum_{s_{ij}\in \{00,11\}}\sum_{s_{\overline{ij}},s_{\overline{ij}}'}
        \langle s_{ij},s_{\overline{ij}}|\rho_0 |s_{ij},s'_{\overline{ij}}\rangle \ket{s_{ij},s_{\overline{ij}}}\bra{s_{ij},s'_{\overline{ij}}}  \\
        &+\sum_{s_{ij}\in \{00,11\}}\sum_{s_{\overline{ij}},s_{\overline{ij}}'}
        \langle s_{ij},s_{\overline{ij}}|\rho_0 X_iX_j  |s_{ij},s'_{\overline{ij}}\rangle \ket{s_{ij},s_{\overline{ij}}}\bra{s_{ij},s'_{\overline{ij}}}X_iX_j  \\
        \equiv & \sum_{s_{ij}\in \{00,11\}}\sum_{s_{\overline{ij}},s_{\overline{ij}}'}\sum_{W\in \{\mathbbm{1}_i \mathbbm{1}_j,X_iX_j\}}
        \langle s_{ij},s_{\overline{ij}}|\rho_0 W  |s_{ij},s'_{\overline{ij}}\rangle \ket{s_{ij},s_{\overline{ij}}}\bra{s_{ij},s'_{\overline{ij}}}W.
    \end{aligned}
\end{equation}

Similarly, for $M_2$ one obtains
\begin{equation}
    \frac{1}{p} M_2\rho_0 M_2^{\dagger} = 
    \sum_{s_{ij}\in \{01,10\}}\sum_{s_{\overline{ij}},s_{\overline{ij}}'}\sum_{W\in \{\mathbbm{1}_i \mathbbm{1}_j,X_iX_j\}}
        \langle s_{ij},s_{\overline{ij}}|\rho_0 W  |s_{ij},s'_{\overline{ij}}\rangle \ket{s_{ij},s_{\overline{ij}}}\bra{s_{ij},s'_{\overline{ij}}}W.
\end{equation}

Combining all contributions, we finally arrive at
\begin{equation}\label{eq:e_ijrho0}
  \mathcal{E}_{\langle ij\rangle}[\rho_0] =  \sum_k M_k\rho M_k = (1-p)\rho_0 + p
    \sum_{s_{ij}}\sum_{s_{\overline{ij}},s_{\overline{ij}}'}\sum_{W}
        \langle s_{ij},s_{\overline{ij}}|\rho_0 W  |s_{ij},s'_{\overline{ij}}\rangle \ket{s_{ij},s_{\overline{ij}}}\bra{s_{ij},s'_{\overline{ij}}}W,
\end{equation}
where $s_{ij}\in\{0,1\}^{\otimes 2}$, $s_{\overline{ij}},s'_{\overline{ij}}\in\{0,1\}^{\otimes (N-2)}$, and
$W\in\{\mathbbm{1}_i \mathbbm{1}_j,X_iX_j\}$. 
Using Eq.~\eqref{eq:rho0_expand}, we rewrite Eq.~\eqref{eq:e_ijrho0} as 

\begin{equation}
  \rho\equiv  \mathcal{E}_{\langle ij\rangle}[\rho_0]  = \sigma_1 + \sigma_2 , 
\end{equation}
where 
\begin{align}
    \sigma_1 
        &\equiv \frac{1-p}{2}
    \sum_{s}\sum_{s'} \sum_{W}
    \langle s|\rho_0 WW |s'\rangle
      \ket{s}\bra{s'}, 
      \label{eq:sigma1_appx}
      \\ 
      \sigma_2  & \equiv 
      p
    \sum_{s_{ij}}\sum_{s_{\overline{ij}},s_{\overline{ij}}'}\sum_{W}
        \langle s_{ij},s_{\overline{ij}}|\rho_0 W  |s_{ij},s'_{\overline{ij}}\rangle \ket{s_{ij},s_{\overline{ij}}}\bra{s_{ij},s'_{\overline{ij}}}W \nonumber \\ 
        & = p
    \sum_{s}\sum_{s_{\overline{ij}}'}\sum_{W}
        \langle s|\rho_0 W  |s_{ij},s'_{\overline{ij}}\rangle \ket{s}\bra{s_{ij},s'_{\overline{ij}}}W.
        \label{eq:sigma2_appx}
\end{align}

\section{Quantum Monte Carlo algorithm}
In this section we describe how to simulate a quantum state $\rho_0$ under a single local decoherence channel $\mathcal{E}_{\langle ij\rangle}$, corresponding to the state in Fig.~\ref{fig:tensor}(b), using QMC, provided that the state $\rho_0$ itself can be simulated by QMC. 
The generalization to the full channel $\mathcal{E}=\prod_{\langle ij\rangle}\mathcal{E}_{\langle ij\rangle}$ in Fig.~\ref{fig:tensor}(e) is straightforward and therefore omitted for brevity.

For illustrations, we take $\rho_0 = e^{-\beta H}$ to be the unnormalized Gibbs state of a Hamiltonian $H$, which can be obtained from finite-temperature QMC simulations. For sufficiently large $\beta$, $\rho_0$ approximates the ground state. More generally, $\rho_0$ may represent a mixed state obtained by tracing out environmental degrees of freedom, and the same implementation of the decoherence channel described below applies as well.

To compute linear observables of $\rho$, we introduce the generalized partition function
\begin{equation} 
    \mathrm{Tr}(\rho) \equiv \mathrm{Tr}(\sigma_1) + \mathrm{Tr}(\sigma_2). 
\end{equation}
Within QMC, the trace is evaluated by sampling configurations in a generalized configuration space. 
Therefore, $\Tr(\rho)$ can be viewed as an extended ensemble composed of two sectors associated with the statistical weights of $\mathrm{Tr}(\sigma_1)$ and $\mathrm{Tr}(\sigma_2)$. By allowing stochastic transitions between these sectors, the simulation dynamically switches between them, so that at any given QMC step, only one of the two contributions needs to be evaluated.

\begin{figure}[ht!]
    \centering
    \includegraphics[width=14cm]{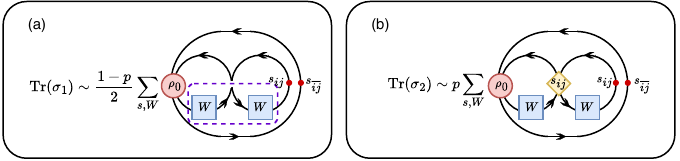}
    \caption{
    Graphical representation of (a) $\Tr(\sigma_1)$ and (b) $\Tr(\sigma_2)$ in the evolution picture, with propagation from the ket to the bra index in the computational basis.
    }
    \label{fig:qmc_transition}
\end{figure}

Figs.~\ref{fig:qmc_transition}(a) and (b) show the evolution pictures (generalized path integral representations) for $\mathrm{Tr}(\sigma_1)$ and $\mathrm{Tr}(\sigma_2)$, respectively. These diagrams are obtained by contracting the bra and ket indices in Figs.~\ref{fig:tensor}(c) and (d), with the time orientation to be counterclockwise without loss of generality.
For a fixed $W$ in each QMC configuration, the spin states propagate along the directed time contour by $W$ operators and the imaginary-time propagation operator $\rho_0$.

Consider a configuration in one of the sectors. Note that the spin variables $s_{ij}$, starting from the initial time slice, always encounter an even number of operators $W\in\{\mathbbm{1}_i \mathbbm{1}_j,X_iX_j\}$ before the application of the imaginary-time propagation operator $\rho_0=e^{-\beta H}$. Consequently, the spin configuration immediately before and after $\rho_0$ remains the same. The QMC updates associated with $\rho_0$ are therefore identical to those used in conventional QMC simulations and can be sampled using standard techniques such as the world-line method (based on the Trotter-Suzuki decomposition)~\cite{Beard1996continuosqmc,Kawashima2004worldline} or the stochastic-series-expansion (SSE, based on a Taylor expansion) method~\cite{sandvik1999sse,Melko2013sse,sandvik2003sseising}.

For updates of the $W$ operators, the two $W$ operators must be updated simultaneously so that they are either both $\mathbbm{1}_i \mathbbm{1}_j$ or both $X_iX_j$, according to Eqs.~\eqref{eq:sigma1} and \eqref{eq:sigma2}. 
For example, within the SSE formulation, this constraint can be incorporated into the Swendsen-Wang-like cluster updates (commonly used for the TFIM, $\mathbb{Z}_2$ lattice gauge theory, Heisenberg model, etc.)~\cite{Swendsen1987,sandvik2003sseising} with the following additional rules:

\begin{itemize} 
    \item \textbf{(Update for $W$ operators.)}
    Before growing a cluster to perform the off-diagonal update, each $W$ operator randomly selects one of the branching choices that determines how the cluster lines enter and exit, as illustrated in Fig.~\ref{fig:branch_ways}.  
    For the two $W$ operators, there are four possible combinations: $(A,A)$, $(A,B)$, $(B,A)$, and $(B,B)$. Since in practice $(A,B)$ and $(B,A)$ are treated in the same way, as shown Fig.~\ref{fig:branch_combo}, only three distinct cases need to be discussed. 
    \begin{figure}[ht!]
        \centering
        \includegraphics[width=10cm]{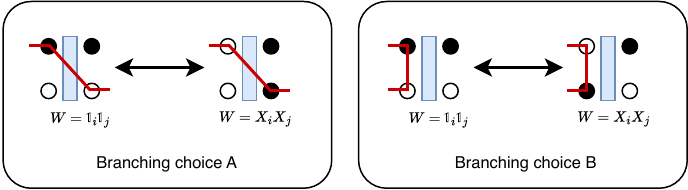}
        \caption{
        Two branching choices when a cluster line (red) encounters a $W$ operator. 
        We follow the standard SSE notation: $\ket{0}$ is represented by an empty circle and $\ket{1}$ by a solid circle. 
        The time direction is horizontal, and the blue square denotes the $W$ operator that propagates the spin states from one side to the other.
        }
        \label{fig:branch_ways}
    \end{figure}

    \begin{figure}[ht!] {
            \centering
            \includegraphics[width=15cm]{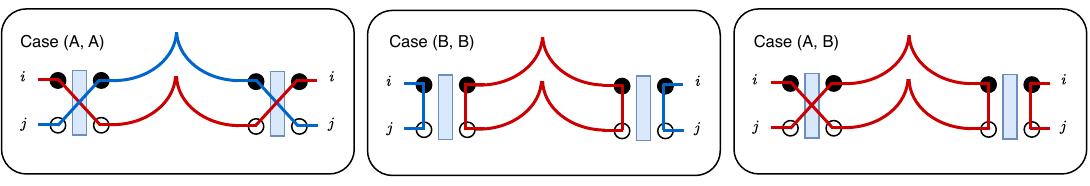}
            \caption{
            Three cases in the cluster update of the $W$ operators on sites $i$ and $j$. 
            The spacetime slice shown corresponds to the region inside the dashed purple square in Fig.~\ref{fig:qmc_transition}(a).
            The red and blue lines denote cluster lines belonging to different clusters.
            }
            \label{fig:branch_combo}
        }
    \end{figure}

    Here we take $\Tr(\sigma_1)$ in Fig.~\ref{fig:qmc_transition}(a) as an example, and one can similarly discuss $\Tr(\sigma_2)$:
    \begin{itemize}
        \item \emph{Case $(A,A)$.} The two $W$ operators belong to at most two independent clusters, whose cluster lines are denoted by the red and blue lines, respectively. Each cluster is flipped with a probability determined by the Metropolis algorithm, by comparing the configuration weights before and after the flip. For the TFIM, for example, this probability is $1/2$.
        \item \emph{Case $(B,B)$.} There are also at most two independent clusters. Note that if a cluster line enters a spin on the leftmost side, it must correspondingly reappear on the rightmost side to ensure the construction of a complete blue cluster.
        \item \emph{Case $(A,B)$.} There is only one cluster. This cluster involves all eight spins associated with the two $W$ operators; therefore, regardless of whether the cluster is flipped, the $W$ operators remain unchanged in this cluster update.
    \end{itemize}
    
    \item \textbf{(Sector update.)}
    To switch between the two sectors, a spin configuration must be compatible with both sectors. In particular, in the $\mathrm{Tr}(\sigma_2)$ sector the Kronecker tensor imposes the constraint that the spins from the four time directions are identical for sites $i$ and $j$. Therefore, if the spins $s_{ij}$ from the four time directions are equal, the sector update proceeds as follows:

    \begin{itemize}
        \item If the configuration lies in the sector of $\mathrm{Tr}(\sigma_1)$, we switch to $\mathrm{Tr}(\sigma_2)$ by inserting the Kronecker tensor that identifies the four time directions at lattice sites $i$ and $j$, with probability
        \begin{equation}
        P(\sigma_1\to \sigma_2)=\min\bigg\{1,\frac{1-p}{2p}\bigg\}.
    \end{equation}
    \item If the configuration lies in the sector of $\mathrm{Tr}(\sigma_2)$, we switch to $\mathrm{Tr}(\sigma_1)$ by removing the Kronecker tensor with probability 
        \begin{equation}
            P(\sigma_2\to \sigma_1)=\min\bigg\{1,\frac{2p}{1-p}\bigg\}.
        \end{equation}
    \end{itemize}
\end{itemize}

The same framework allows the simulation of $\mathrm{Tr}(\rho^2)$, enabling the evaluation of nonlinear two-copy observables such as the Rényi-2 correlator $C^{(2)}$ and the Rényi-2 linear order correlator $C^{(1)}$ in the doubled space. This corresponds to sampling an extended ensemble with four sectors
\begin{equation}
    \mathrm{Tr}(\rho^2)
    =
    \mathrm{Tr}(\sigma_1^2)
    +
    \mathrm{Tr}(\sigma_1\sigma_2)
    +
    \mathrm{Tr}(\sigma_2\sigma_1)
    +
    \mathrm{Tr}(\sigma_2^2).
\end{equation}

The associated path-integral representations follow analogously, as shown in Fig.~\ref{fig:tensor}(f), and the QMC updates are implemented in the same manner. Because the computational basis is chosen to be the $Z$ basis, both $C^{(1)}$ and $C^{(2)}$ correspond to diagonal measurements when sampling $\mathrm{Tr}(\rho^2)$, and can therefore be evaluated with efficiency comparable to conventional QMC simulations.

In closing, we discuss the computational complexity of the QMC scheme introduced here.
In our generalized path-integral formulation, the effective imaginary-time extent receives two contributions: the physical inverse temperature $\beta$ and an additional term proportional to the number of bonds $N_b = \Theta(L^{d})$ arising from $W$-operator insertions, where $d$ is the spatial dimension. As a result, the total imaginary-time length scales as $\Theta(\max\{\beta, L^d\})$.
Consequently, the cost of a single Monte Carlo sweep is proportional to the spacetime volume (or configuration size) $\mathcal{V} \sim \mathcal{O}\!\big(\max\{\beta, L^d\}\,L^d\big)$.
Therefore, the overall computational complexity follows the same scaling as in standard QMC approaches, with $\mathcal{V}$ replacing the usual spacetime volume, and remains polynomial in system size and inverse square of the error tolerance. 
Correspondingly, evaluating Rényi-2 observables requires only a constant (factor of two) overhead compared to conventional linear observables, as the simulation involves two replicas of the system.

\section{Derivation of the Effective 2D Defect Action}

In this section, we explicitly integrate out the massive bulk degrees of freedom to derive the effective two-dimensional action on the defect. We start with the bulk action for a single replica $\phi$ (omitting the replica index $a,b$ for brevity), treating the $\lambda \phi^4$ term as a perturbation. The quadratic part of the bulk action is
\begin{equation}
    S_0[\phi] = \int d^2x dz \left[ \frac{1}{2}\partial_i \phi \partial^i \phi + \frac{1}{2}(\partial_z \phi)^2 + \frac{1}{2}m^2 \phi^2 \right],
\end{equation}
where $x^i$ ($x^i$=x,y) denotes the coordinates strictly parallel to the defect , and $z$ is the perpendicular coordinate. The defect is located at $z=0$.

To integrate out the bulk at tree level, we solve the classical equation of motion (EOM) for the bulk field in the region $z \neq 0$:
\begin{equation}
    (-\partial_i \partial^i - \partial_z^2 + m^2) \phi(x_i, z) = 0.
\end{equation}
Performing a Fourier transform along the defect directions, $\phi(x_i, z) = \int \frac{d^2k}{(2\pi)^2} e^{i k^i x_i} \tilde{\phi}(k^i, z)$, the EOM decouples for each momentum mode $k^\mu$:
\begin{equation}
    (-\partial_z^2 + k^2 + m^2) \tilde{\phi}(k^i, z) = 0,
\end{equation}
where $k^2 = k_i k^i$. Requiring the field to decay at $z \to \pm\infty$, the solution is given by separation of variables:
\begin{equation}
    \tilde{\phi}(k^i, z) = \varphi(k^i) e^{-\sqrt{k^2+m^2}|z|},
\end{equation}
where $\varphi(k^i) \equiv \tilde{\phi}(k^i, z=0)$ is the boundary field strictly confined to the defect.

To evaluate the effective action for the defect field, we substitute the classical solution back into the bare quadratic action $S_0$. Using integration by parts, the action can be rewritten as
\begin{equation}
    S_0[\phi] = \frac{1}{2} \int d^2x \left[ \int_{-\infty}^{0^-} dz + \int_{0^+}^{\infty} dz \right] \left[ \partial_i(\phi \partial^i \phi) + \partial_z(\phi \partial_z \phi) - \phi(\partial_i \partial^i \phi + \partial_z^2 \phi - m^2 \phi^2) \right].
\end{equation}
Since the classical profile strictly satisfies the EOM $(-\partial_i \partial^i - \partial_z^2 + m^2)\phi = 0$ in the bulk ($z \neq 0$), the last term in the integrand vanishes identically. Assuming the field decays to zero at $z \to \pm \infty$, the integration over $z$ leaves only the boundary terms at the defect $z=0$:
\begin{equation}
    S_0[\phi] = \frac{1}{2} \int d^2x \left[ \phi(x_i, 0) \partial_z \phi(x_i, 0^-) - \phi(x_i, 0) \partial_z \phi(x_i, 0^+) \right].
\end{equation}
Transforming to momentum space along the defect directions, we use the classical profile $\tilde{\phi}(k^i, z) = \varphi(k^i) e^{-\sqrt{k^2+m^2}|z|}$. The normal derivatives at the two sides of the defect are
\begin{align}
    \partial_z \tilde{\phi}(k^i, 0^+) &= -\sqrt{k^2+m^2} \varphi(k^i), \\
    \partial_z \tilde{\phi}(k^i, 0^-) &= +\sqrt{k^2+m^2} \varphi(k^i).
\end{align}
Substituting these derivatives back into the boundary action, the discontinuity of the derivative across $z=0$ yields a factor of $2\sqrt{k^2+m^2}$, which precisely cancels the $1/2$ prefactor. Thus, we obtain the effective quadratic action for the boundary field $\varphi$:
\begin{equation}
    S_{\text{eff}}^{(0)}[\varphi] = \int \frac{d^2k}{(2\pi)^2} \varphi(-k^i) \sqrt{k^2+m^2} \varphi(k^i).
\end{equation}

In the trivial bulk phase where $m > 0$, the bulk correlation length $\xi = 1/m$ is finite. For low-energy fluctuations on the defect ($k \ll m$), we can expand the non-local kinetic kernel as $\sqrt{k^2+m^2} \approx m + \frac{k^2}{2m} + \mathcal{O}(k^4)$. Transforming back to real space, the effective quadratic action becomes local:
\begin{equation}
    S_{\text{eff}}^{(0)}[\varphi] \approx \int d^2x \left[ \frac{1}{2m} \partial_i \varphi \partial^i \varphi + m \varphi^2 \right].
\end{equation}

To account for the bulk interactions, we evaluate the tree-level contribution of the bulk $\lambda \phi^4$ term using the low-energy classical profile $\phi(x^i, z) \approx \varphi(x^i) e^{-m|z|}$. This approximation is equivalent to neglecting derivative interactions generated by the integration over the bulk, which are strictly irrelevant in the renormalization group sense and do not affect the low-energy universal behavior.
 Integrating over the perpendicular coordinate $z$ yields
\begin{equation}
    \int d^2x \int_{-\infty}^{\infty} dz \, \lambda \left( \varphi(x^i) e^{-m|z|} \right)^4 = \int d^2x \frac{\lambda}{2m} \varphi^4(x^i).
\end{equation}
Physically, the factor of $1/m$ reflects the bulk correlation length $\xi$, indicating that only the bulk degrees of freedom within a distance $\xi$ from the defect effectively participate in the interaction.

Here we derive the effective boundary action $S_{\text{int}}[\phi_a,\phi_b]$ from the microscopic interaction $V_{\text{int}} = 2 \delta(z)u \sum_{\langle i,j\rangle} Z_i^a Z_j^a Z_i^b Z_j^b$, where $u \equiv \tanh^{-1}[p/(2-p)]$.

In the continuum limit, the lattice spin operator $Z_i$ maps to the scalar field $\phi(x)$. Using the operator product expansion (OPE), the nearest-neighbor spin product can be expanded in terms of the identity and the energy density operator $\epsilon(x) \propto \phi^2(x)$. Thus, we write the coarse-grained bond operator as $Z_i Z_j \simeq C_0 + C_1 \phi^2(x)$, with $C_{0,1}$ being non-universal constants. Applying this to both replicas $a$ and $b$, the four-spin interaction becomes:
\begin{equation}
    Z_i^a Z_j^a Z_i^b Z_j^b \simeq C_0^2 + C_0 C_1 \left( \phi_a^2 + \phi_b^2 \right) + C_1^2 \phi_a^2 \phi_b^2.
\end{equation}

Taking the continuum limit $\sum_{\langle i,j\rangle} \to \frac{1}{a^2}\int d^2x$ and dropping the trivial constant shift $C_0^2$, the continuous interaction action $S_{\text{int}} = -\int dz V_{\text{int}}$ takes the form:
\begin{equation}
    S_{\text{int}} \simeq -\frac{2u}{a^2} \int d^2x dz \delta(z) \left[ C_0 C_1 \left( \phi_a^2 + \phi_b^2 \right) + C_1^2 \phi_a^2 \phi_b^2 \right].
\end{equation}

Comparing this with the phenomenological 2D Ashkin-Teller boundary action,
\begin{equation}
    S_{\text{int}}[\phi_a,\phi_b] = -\int d^2x dz \delta(z) \left[ \tilde{m}\left(\phi_a^2+\phi_b^2\right) + t\phi_a^2\phi_b^2 \right],
\end{equation}
we can explicitly match the coefficients to find $\tilde{m} = 2 C_0 C_1 u$ and $t = 2 C_1^2 u$.

To bring the effective action into the standard canonical form, we rescale the boundary field as $\varphi \longrightarrow \sqrt{m} {\varphi}$, which normalizes the kinetic term to $\frac{1}{2}\partial_i {\varphi} \partial^i {\varphi}$. Under this rescaling, the effective quartic coupling becomes $\tilde{\lambda} = \lambda m / 2$. Restoring the replica indices $a, b$, we arrive at the full effective 2D action:
\begin{equation}
    S_{\text{eff}}[{\varphi}^a, {\varphi}^b] = \int d^2x \left[ \sum_{\alpha=a,b} \frac{1}{2}\left( \partial_i {\varphi}_\alpha \partial^i {\varphi}_\alpha + m_{\text{eff}}^2 \varphi_\alpha^2 + \tilde{\lambda} \varphi_\alpha^4 \right) - \tilde{t} {\varphi}_a^2{\varphi}_b^2 \right],
\end{equation}
where $m_{\text{eff}}^2 =m(2m - \tilde{m})$ incorporates the bare mass and the defect potential $\tilde{m}$, $\tilde{t}=tm^2$. This is exactly the continuum field theory of the 2D Ashkin-Teller model.

Based on the effective 2D Ashkin-Teller action derived above, we can understand the phase diagram using Landau Ginzburg mean-field theory. The exact phase diagram can be found in \cite{Aoun:2024lmd}, where, upon a 90° rotation, it matches our results precisely. Assuming uniform field configurations, the Landau free energy density is given by the potential terms of the effective action:
\begin{equation}
    f(\varphi_a, \varphi_b) = \frac{1}{2}m_{\text{eff}}^2 (\varphi_a^2 + \varphi_b^2) + \frac{1}{2}\tilde{\lambda} (\varphi_a^4 + \varphi_b^4) - \tilde{t} \varphi_a^2 \varphi_b^2 +O(\varphi_a^3\varphi_b^3),
\end{equation}

The ground state is determined by minimizing the free energy density. The extremum conditions $\partial f / \partial \varphi_a = 0$ and $\partial f / \partial \varphi_b = 0$ yield the equations of state:
\begin{align}
    \varphi_a (m^2_{\text{eff}} + 2\tilde{\lambda} \varphi_a^2 - 2\tilde{t} \varphi_b^2) &= 0, \\
    \varphi_b (m^2_{\text{eff}} + 2\tilde{\lambda} \varphi_b^2 - 2\tilde{t} \varphi_a^2) &= 0.
\end{align}

By solving these equations, we identify three distinct physical regimes, which perfectly correspond to the phases observed in our QMC simulations:

\textbf{1. Trivial Symmetric Phase:}
When $m^2_{\text{eff}}>0$ and $\tilde{\lambda}>\tilde{t}$  , the only real solution is the trivial vacuum:
\begin{equation}
    \langle \varphi_a \rangle = \langle \varphi_b \rangle = 0.
\end{equation}
 In this regime, the system is in the strongly symmetric phase, and no symmetry is broken. This corresponds to the region where both the TFIM $J$ and the decoherence strength $p$ are small.

\textbf{2. R2-SSB Phase:}
When $m^2_{\text{eff}} < 0$ , the trivial vacuum becomes unstable. If the inter-replica coupling $t$ is positive and relatively small, the system prefers to develop nonzero vacuum expectation values (VEVs) for both replicas simultaneously to minimize the $-t \varphi_a^2 \varphi_b^2$ interaction. The minimum occurs at:
\begin{equation}
    \langle \varphi_a \rangle^2 =\langle \varphi_b \rangle^2 = \frac{-m^2_{\text{eff}}}{2\tilde{\lambda} - 2\tilde{t}}= \frac{\tilde{m}-2m{}}{{\lambda} - 2{t}m}.
\end{equation}

In this phase, both $\langle \varphi_a \rangle \neq 0$ and $\langle \varphi_b \rangle \neq 0$, which implies that the $\mathbb{Z}^a_2 \times \mathbb{Z}^b_2$ symmetry is completely broken down to the trivial group. Both $C^{(1)}$ and $C^{(2)}$ are nonzero. 

Crucially, the ferromagnetic phase is stable only when the inter-replica coupling satisfies $t <  \lambda/2m$. Near the bulk critical point ($J \to J_c$), the small mass gap $m$ yields a large $t$, stabilizing this phase over a wide range of $p$. Conversely, deeper in the trivial phase (large $m$), achieving negative $m^2_{\text{eff}}$ requires a large decoherence strength $p$. Since $t$ grows monotonically with $p$, it inevitably exceeds $\lambda$, violating the stability condition. Consequently, for sufficiently small $J$, the ferromagnetic phase is entirely preempted, and the system transitions directly into the Baxter phase.

\textbf{3. R2-SWSSB phase:}
As the decoherence strength $p$ increases, the inter-replica coupling $t$ becomes sufficiently large. In the Ashkin-Teller model, a large $t$ renders the ferromagnetic phase unstable, as it strongly binds the fluctuations of the two replicas. Consequently, the naive Landau-Ginzburg description is not positive definite and hence becomes inadequate. To properly capture the physics in this regime, one must include higher-order interaction terms  and consider the composite field $\sigma \sim \varphi_a \varphi_b$ as the relevant order parameter. In this phase, the strong inter-replica coupling prevents the individual fields from acquiring a vacuum expectation value ($\langle \varphi_a \rangle = \langle \varphi_b \rangle = 0$), but their composite operator acquires a nonzero expectation value:
\begin{equation}
    \langle \varphi_a \varphi_b \rangle \neq 0.
\end{equation}
This composite order parameter breaks the full $\mathbb{Z}^a_2 \times \mathbb{Z}^b_2$ symmetry but strictly preserves the diagonal subgroup $\mathbb{Z}_2^{\text{diag}}$ (under which $\varphi_a \to -\varphi_a$ and $\varphi_b \to -\varphi_b$ simultaneously). Consequently, the linear correlator vanishes ($C^{(1)} = 0$), while $C^{(2)} \neq 0$, marking the hallmark of the SWSSB phase.

\end{document}